\documentclass[aps,prd,twocolumn,superscriptaddress,nofootinbib]{revtex4-2}

\pdfoutput=1

\usepackage{amsmath}
\usepackage{dsfont}
\usepackage{graphicx}
\usepackage{float}
\usepackage{bbold} 
\usepackage{xspace} 
\usepackage[caption=false]{subfig} 
\usepackage{grffile}
\usepackage{slashed}
\usepackage{rotating,multirow} 

\usepackage{simplewick} 
\usepackage[prependcaption,textsize=tiny]{todonotes}

\newcommand \spinhalf {spin-$\frac{1}{2}$\xspace}
\newcommand \mpcac {m_{\mathrm{PCAC}}}

\newcommand{\tr}{\mathrm{T}}
\newcommand{\tra}{\mathrm{tr}}
\newcommand{\Tr}{\text{Tr}}
\newcommand\pmp{\xi_{+}}
\newcommand\pmm{\xi_{-}}
\newcommand{\xib}{\bar{\xi}}

\newcommand\nubar{\overline{\nu}}
\newcommand\Sgroup[2]{\mathrm{#1}(#2)}
\newcommand\su[1]{\ensuremath{\Sgroup{SU}{#1}}}
\newcommand\so[1]{\ensuremath{\Sgroup{SO}{#1}}}
\newcommand\uone{\ensuremath{\Sgroup{U}{1}}}

\newcommand{\OFAQCD}{SU2Nf1Adj}

\newcommand\Nx[1]{N_{\textnormal{#1}}}
\newcommand\Nf{\Nx{f}}

\begin{document}


\title{Investigating the conformal behaviour of $\su{2}$ with one adjoint Dirac flavor}


\author{Andreas Athenodorou}
\email{andreas.athinodorou@pi.infn.it}
\affiliation{Dipartimento di Fisica, Universit\`a di Pisa and INFN, Sezione di Pisa, Largo Pontecorvo 3, 56127 Pisa, Italy.}

\author{Ed Bennett}
\email{e.j.bennett@swansea.ac.uk}
\affiliation{Swansea Academy of Advanced Computing, Swansea University, Fabian Way, Swansea SA1 8EN, UK}

\author{Georg Bergner}
\email{georg.bergner@uni-jena.de}
\affiliation{University of Jena, Institute for Theoretical Physics, Max-Wien-Platz 1, D-07743 Jena, Germany}

\author{Biagio Lucini}
\email{b.lucini@swansea.ac.uk}
\affiliation{Department of Mathematics, Swansea University, Fabian Way, Swansea SA1 8EN, UK}


\date{September 10, 2021}

\begin{abstract}
  We present a major update on our investigations of $\su{2}$ gauge theory with one Dirac flavor in the adjoint representation on the lattice. In particular we consider larger volumes, as well as four different values of the gauge coupling. We provide results for the spectrum including gluonic, fermionic, and hybrid observables, Polyakov loops, and the anomalous dimension of the fermionic condensate from the Dirac mode number. 
  These data confirm that the theory is close to the lower boundary of the conformal window for adjoint fermions. Our investigations provide important insights regarding the realization of different infrared scenarios that have been conjectured for this theory.
\end{abstract}

\pacs{11.15.Ha \and 12.60.Nz}

\maketitle


\section{Introduction}
\label{sect:introduction}
The \su{2} gauge theory with $N_f$ Dirac fermions in the adjoint representation has seen many interesting recent applications in theoretical 
studies ranging from composite Higgs models to topological phases related 
to systems of condensed matter physics. The most prominent examples of these theories are
the $N_f=2$ case, called Minimal Walking Technicolor, and supersymmetric Yang-Mills theory, corresponding to the case with one Majorana fermion. Recently also the case of $N_f=1$ Dirac fermion gained much interest. It has been considered in the context of semiclassical analysis \cite{Unsal:2007jx} and volume independence leading to the conjecture of emergent fermion symmetry event in the non-supersymmetric case \cite{Basar:2013sza}. Another line of interest has been the relation of chiral symmetry breaking and confinement. The  most recent interest has come from the consideration of 't Hooft anomaly matching and topological phase transitions \cite{Bi:2018xvr,Anber:2018tcj}. Although all of these investigations reveal interesting aspects and conjectures about the non-perturbative regime, it is still unclear what kind of phase is realized at low temperatures. It should be stressed that all conjectures and studies depend on this basic knowledge of the theory.

The main motivation of our previous investigations of the \su{2} gauge theory with one adjoint Dirac fermion has been the question about signatures of a conformal window related to composite Higgs scenarios.
In these scenarios the Higgs sector emerges as a low energy manifestation of novel strong
dynamics~\cite{Weinberg:1975gm,Susskind:1978ms,Eichten:1979ah,Holdom:1981rm,Holdom:1984sk}.
This new strong interaction is able to 
explain the observed electroweak symmetry breaking phenomenology if
the following three conditions are met: (1)
the theory is near the onset of the conformal window; (2)
the anomalous dimension of the chiral condensate is of
order one; and (3) a parametrically light scalar (the would-be Higgs
boson) is in the spectrum. The first two
conditions~\cite{Yamawaki:1985zg,Appelquist:1986an} are needed for
compatibility with electroweak precision data~\cite{Peskin:1991sw}, while the third
condition is determined by the direct observation of the Higgs boson
and no other previously unknown nearby state. Numerical lattice simulations have 
provided important information about the realization of these conditions in different 
theories. Interesting candidates are in particular gauge theories coupled to fermions in 
higher representations since these require a smaller number of fermion fields to reach the conformal window~\cite{Sannino:2003xe,Sannino:2004qp,Dietrich:2005jn,Dietrich:2006cm}.
One example of a gauge theory which, according to numerical data, is inside the conformal window is $\su{2}$ gauge theory
with two adjoint Dirac
flavors~\cite{DelDebbio:2009fd,Hietanen:2009az,Catterall:2009sb,Bursa:2009we,DelDebbio:2010hu,DelDebbio:2010hx,Bursa:2011ru,Catterall:2011zf,DeGrand:2011qd,Catterall:2007yx,DelDebbio:2008zf,Hietanen:2008mr,Catterall:2008qk}. However, the low mass anomalous dimension\footnote{The actual mass anomalous dimension might even be lower, see Ref.~\cite{Bergner:2016hip} and references therein.}, $\gamma_{\star} = 0.38(2)$~\cite{DelDebbio:2013hha}, at the infrared fixed point disfavors this model for phenomenological applications. Meanwhile also a first study with three Majorana fermions in the adjoint representation has appeared \cite{Bergner:2017gzw} showing an increasing mass anomalous dimension towards a lower fermion content of the theory.

The identification of a theory with an at least near-conformal behavior and a large mass anomalous dimension has been the main motivation for our previous studies of \su{2} with a single adjoint Dirac flavor. In fact, given that the theory with two adjoint flavors was observed to be well within the conformal window, there is the possibility that the theory with a single adjoint flavor sits at the onset of conformality. Such behavior is required for phenomenologically viable theories to explain electroweak symmetry breaking at a fundamental level. In this context, a first obvious but very relevant question is whether the theory is just outside or inside the conformal window. The theory can later be extended by additional fundamental fermions to provide a viable extension of the Standard Model \cite{Ryttov:2008xe,Bergner:2020mwl}.

Close to the onset of the conformal window, it is hard to distinguish a conformal from a chiral symmetry breaking scenario.\footnote{We observe that for fermions in the adjoint representation the chiral symmetry breaking scale can be separated from the confinement scale; see~\cite{Evans:2020ztq} for a recent discussion} In such a (near-)conformal regime, large scaling corrections are also expected. Therefore a very detailed study of the dependence on simulation parameters is required. In this respect, our first investigations \cite{Athenodorou:2014eua} at only a single lattice spacing have provided only a limited insight. A later study in \cite{Bi:2019gle} considered mainly the same parameter range for a different analysis. For this reason, we provide here a substantial extension of our previous simulations considering a range of different gauge couplings and a large number of additional ensembles in order to improve on our first investigations of the theory.

Another line of motivation for an investigation of $\su{2}$ with one adjoint Dirac flavor is a general scan in the parameter space of strongly interacting theories to identify common patterns and relate them to analytical predictions, for example based on gauge/gravity duality. In this context we have investigated the ratio of lightest spin-2 and spin-0 resonances \cite{Athenodorou:2016ndx}. We have shown that possible non-trivial universal characteristic in the landscape of strongly interacting gauge theories can be identified. These could later on help to find candidate theories for realistic standard model extensions.

An interesting further relation of the theory is with ${\cal N} = 2$ super Yang--Mills with gauge group
$\su{2}$, in the limit of a large scalar mass which leads to complete supersymmetry breaking.  In ${\cal N} = 2$ super Yang--Mills, confinement is
known to arise through the dual superconductor mechanism resulting from magnetic monopole condensation~\cite{Seiberg:1994rs}. The breaking of supersymmetry by a finite scalar mass has been considered in~\cite{Cordova:2018acb}, where the authors argue that in the large mass limit chiral symmetry breaking is at play. 

Our investigations are based on the first-principles method of numerical Monte Carlo simulations on a spacetime lattice. 
In Section \ref{sect:model} we provide a summary of the continuum properties, and lattice action, of the theory. We also explain the main observables and parameters considered in this study. Additional information about the simulation setup and the lattice formulation can be found in our previous publication \cite{Athenodorou:2014eua}.

The main results presented in Section \ref{sect:results} start with a discussion on the general limitations of the parameter range due to phase transitions and topological freezing. We show that our simulations are not limited by these general systematic effects. The obtained bound state particle spectrum at different fermion masses is presented in Section \ref{sec:particle}. The main result of this paper regarding mass anomalous dimension are presented in Section \ref{sec:anomalous}. Assuming conformal scaling, we determine the mass anomalous dimension from particle spectrum and mode number. The results confirm a significantly larger anomalous dimension than in the two Dirac flavor case. However, they also reveal relevant scaling corrections. We discuss as an alternative scenario also a fit according to chiral perturbation theory in Section \ref{sec:chipt}. As a short update regarding our work in \cite{Athenodorou:2016ndx}, we also discuss the ratio between the spin-2 and spin-0 glueballs. At the end of the paper we discuss the implications for infrared scenarios of the theory in Section \ref{sec:con}.

\section{Summary of the theory and considered observables}
\label{sect:model}
We investigate the \su{2} gauge theory with a single Dirac fermion in the adjoint representation (\OFAQCD{}). Although our goal is to understand the theory in the massless limit, numerical simulations require a finite fermion mass. We, therefore, simulate at different fermion masses and investigate how the massless limit is approached. Our primary goal is to understand whether chiral symmetry is spontaneously broken in the massless limit or the theory is infrared (IR) conformal. In case of spontaneous chiral symmetry breaking, chiral perturbation theory would be sufficient to describe the low energy effective theory at small fermion masses. On the contrary, if our theory is IR conformal, the data will be in accordance with the scaling predictions of a mass-deformed IR conformal gauge theory. It is possible that \OFAQCD{} is in the QCD-like phase with confinement and chiral symmetry breaking but close to the onset of the conformal window. If such a scenario holds, the system might demonstrate mass-deformed conformal behaviour within an intermediate energy regime between the chiral symmetry breaking scale $\Lambda_{\rm IR}$ and the perturbative scale in ultraviolet $\Lambda_{\rm UV}$, while for energies below $\Lambda_{\rm IR}$ the system will be correctly described by chiral perturbation theory. The latter possibility refers to the so called near-conformal or walking scenario. We focus mainly on the distinction of chiral symmetry breaking and IR conformal scenario, but we also discuss other possible IR scenarios in the final conclusions.

The Lagrangian of \OFAQCD{} in Minkowski space, is defined as:
\begin{eqnarray}
\label{eq:actcont}
{\cal L} = \overline{\psi}(x) \left( i \slashed{D} - m \right) \psi(x)
- \frac{1}{2} \mathrm{Tr} \left( G_{\mu \nu}(x) G^{\mu \nu} (x) \right)\,,
\end{eqnarray}
with $\slashed{D} = \left(\partial_{\mu} + i g A_{\mu}(x) \right)
	\gamma^{\mu}$, $\gamma_{\mu}$ being the Dirac matrices, $A_{\mu}(x) =
	\sum_a T^a A_{\mu}^{a}(x)$ with $a = 1,2,3$, and the $T_a$ the Lie algebra
	generators of $\su{2}$ in the adjoint representation. The field strength tensor is defined as $G_{\mu \nu} = \partial _{\mu} A_{\nu}(x) - \partial
	_{\nu} A_{\mu}(x) + i g [A_{\mu}(x), A_{\nu}(x)]$, with $g$ being the
	gauge coupling of the theory, and the trace is taken
	over the gauge group. Notation and conventions are explained in \cite{Athenodorou:2014eua}.

The adjoint representation is real and, as a consequence, does not mix real and imaginary parts of the Dirac spinor. This enables us to decompose the Dirac spinor in Majorana components as follows:
\begin{eqnarray}
\pmp = \frac{\psi + C \overline{\psi}^\tr}{\sqrt{2}} \ ,
\qquad
\pmm = \frac{\psi - C \overline{\psi}^\tr}{\sqrt{2}i} \ ,
\end{eqnarray}
such that
\begin{eqnarray}
\psi =\frac{1}{\sqrt{2}}( \pmp + i \pmm) \ .
\end{eqnarray}
In the above expressions $C$ is  the  charge  conjugation matrix; both $\pmp$ and $\pmm$ are invariant under charge
conjugation symmetry by construction.  Eq.~(\ref{eq:actcont}) can, thus, be reformulated as
\begin{eqnarray}
{\small
\label{eq:actcont2}
{\cal L} =\frac{1}{2} \hspace{-0.05cm} \sum_k \hspace{-0.05cm} \xib_k(x) \hspace{-0.05cm} \left( i \slashed{D} \hspace{-0.05cm}  -  \hspace{-0.05cm} m \right) \xi_k(x)
\hspace{-0.1cm} - \hspace{-0.1cm}  \frac{1}{2} \mathrm{Tr} \left( G_{\mu \nu}(x) G^{\mu \nu} (x)
\right)}
\end{eqnarray}
where $k = +, -$. This flavor structure (in terms of the Majorana
components) gives rise to a  non-trivial chiral symmetry breaking pattern.

The general chiral symmetry breaking pattern that can be identified from this formulation is
\begin{equation}
\su{2 \Nf} \mapsto \so{2 \Nf} \ .
\end{equation}
In the present case of $\Nf=1$ there are hence two Goldstone
bosons in this model if chiral symmetry is spontaneously broken.
The unbroken $\so{2}$ is equivalent to $\uone$ baryon number in Dirac fermion formulation.

The identification of quantum numbers for two-fermion operators has been presented in our earlier paper \cite{Athenodorou:2014eua}.
These are in principle generalization of the meson operators in QCD, but the labeling is done according the the unbroken U(1).
Hence there are mesons transforming trivially under this symmetry and baryons (or diquark) operators that carry baryon number $\pm 2$.
The operators for the Goldstone bosons correspond to scalar baryons for which the fermion contractions are similar to the pions in QCD. In these extended measurements we don't consider operators with disconnected contributions. Besides the scalar baryons ($\psi^TC\gamma_5\psi$) we consider also the pseudoscalar baryon ($\psi^T C \psi$), the vector baryon ($\psi^TC\gamma_k\gamma_5\psi$), and the vector meson ($\overline{\psi}\gamma_k\psi$).

 \subsection{Lattice action and observables}
Our simulations are based on a lattice action consisting of a Wilson gauge and fermion action given by
\begin{eqnarray}
S=S_{\mathrm{G}} + S_{\mathrm{F}} \,,
\end{eqnarray}
where 
\begin{eqnarray}
  S_{\mathrm{G}} = \beta \sum_{p} {\rm Tr} \left[ 1 - \frac{1}{2}U(p)  \right] \,,
\end{eqnarray}
and 
\begin{eqnarray}
S_{\mathrm{F}} =  \sum_{x,y} {\overline \psi}(x) D(x,y) \psi(y)\,.
\end{eqnarray}
Here $U(p)$ is the lattice plaquette and  $S_{\mathrm{G}}$, $S_{\mathrm{F}}$ are respectively the pure gauge part and the fermionic contribution in the action. The massive Dirac operator is defined as 
\begin{align}
  D(x,y) = \delta_{x,y} 
  - \kappa &\left[ \left(1-\gamma_{\mu}  \right) U_{\mu} (x) \delta_{y,x+\mu} \right.\\
    \nonumber 
    &+  \left. \left(1+\gamma_{\mu}  \right) U^{\dagger}_{\mu} (x-\mu) \delta_{y,x-\mu}   \right]\,,
\end{align}
where $U(p)$ is the lattice plaquette, $\kappa = 1/(8+2 a m)$ is the hopping parameter, $a$ the lattice spacing, and $m$ the bare fermion mass. Further details of the lattice representation and algorithms can be found 
in our earlier work \cite{Athenodorou:2014eua}. As detailed in this reference, particle masses are obtained from correlators of operators projecting onto the selected quantum numbers. Note that the Wilson fermions break chiral symmetry and require the inclusion of an additive and multiplicative renormalizaton of the fermion mass. The \emph{partially conserved axial current} (PCAC) defines a mass $\mpcac$ which requires only multiplicative renormalization which is used as a proxy of the renormalized fermion mass. The lattice
technology used to define correlators and the $\mpcac$
  mass and to compute them on the lattice is by now standard (see
  e.g.~\cite{DelDebbio:2007wk} for a more extended treatment).

For the mesonic observables and the generation of the configurations we have used HiRep\footnote{See https://github.com/claudiopica/HiRep for the HiRep source code.}~\cite{DelDebbio:2008zf,DelDebbio:2009fd}. The results were 
also cross-checked with a code developed for simulations of Super Yang--Mills
theories~\cite{Bergner:2013nwa}. The latter code has also been used to measure a
\spinhalf {\color{black} hybrid fermion} state which is constructed in the continuum from the operator 
\begin{equation}
O_{\textnormal{spin-}\frac{1}{2}} = \sum_{\mu,\nu} \sigma_{\mu\nu} \tra [F^{\mu\nu} \xi]\;,
\end{equation}
where $\sigma_{\mu\nu} = \frac{1}{2}[\gamma_\mu, \gamma_\nu]$. This state,
which can be seen as a bound state of a fermion and a gluon, was
computed using the tools described in~\cite{Bergner:2013nwa}.
Gluonic observables (and in particular glueball and  torelon states) have been studied using the techniques
exposed in~\cite{Lucini:2010nv}. Further details are explained in sections presenting the numerical results{\color{black}, and raw data are available in Ref.~\cite{datapackage}}.

In addition we also determine the scale setting quantities $t_0$ and $w_0$ from the gradient flow. These are introduced in Section \ref{sec:topology}. We also consider the string tension with the methods already explained in~\cite{Athenodorou:2014eua}.

The theory requires a careful consideration of the simulation parameter range. A detailed investigation over the parameter space in~\cite{Athenodorou:2014eua} shows a strong signal for a bulk phase transition at around $\beta \approx 1.9$, $am \approx -1.65$. The requirement to connect with continuum limit, therefore, imposes to simulate at $\beta > 1.95$. In order to avoid the bulk phase and retrieve the desired continuum physics, in our previous study we have chosen the value $\beta=2.05$.
According to this analysis, this value of the gauge coupling is in the region connected to continuum physics. However, it is still quite close to the bulk phase and, assuming a positive $\beta$-function,  corresponds to a rather coarse lattice.  The choice was dictated by the necessity to study large volumes in order to explore the infrared behaviour of the theory while keeping under control the computational cost. However, a thorough assessment of discretisation effects requires studies over a range of increasing $\beta$s, in order to extrapolate to the continuum limit. For these simulations, a higher computational effort is needed. Another reason to simulate at weaker couplings is related to the generic continuum structure of an IR conformal gauge theory, which, in addition to the asymptotically free phase connected with the ultraviolet Gaussian fixed point, is expected to have a strong coupling phase in which the coupling decreases at large distances~\cite{Giedt:2011kz}. Therefore, if the lattice model is investigated at strong coupling, there is the potential danger that in the continuum it will describe the latter phase of the model, which is not the one we are interested in. 

Simulations of a range of larger gauge couplings require a consideration of additional relevant effects that can affect the results. Due to the decreased lattice spacing, the physical lattice volume could become small enough to induce a transition to a phase with broken center symmetry in some direction indicated by a non-zero expectation value of spatial or temporal Polyakov loops. Another important effect is topological freezing. At larger values of~$\beta$ the topological charge fluctuations are strongly suppressed and hence the simulation becomes non-ergodic. We have performed a detailed consideration of these effects in Sections~\ref{sec:center_symmetry} and \ref{sec:topology}.

Our investigation spans over four values of $\beta $, namely $\beta=2.05$, $2.10$, $2.15$ and $2.20$, as well as over a broad range of bare masses $am$. The details of all the ensembles produced for the purposes of this work are presented in Appendix~\ref{sec:lattice_parameters}{\color{black}, and are available in machine-readable form in Ref.~\cite{datapackage}}. For simplicity we denote our ensembles as DB$i$M$j$ where the first index $i=1, \ldots, 4$ refers to the four values of $\beta$ while the second index $j=1, \ldots$  refers to the choice of mass. {\color{black} Furthermore, we use asterisks as subscripts on this notation to denote smaller volumes produced  for investigating finite volume effects and possible phase transitions.} 

\section{Results}
\label{sect:results}
\subsection{Expectation value of Polyakov loops---Center Symmetry}
\label{sec:center_symmetry}
	The introduction of adjoint fermions in \su{N} gauge theories, at zero temperature   and   infinite   spatial   volume preserves  the $\left( \mathds{Z}(N) \right)^4$ symmetry related to center transformations in the four Euclidean directions. When decreasing the volume or increasing the temperature, the system can go through various phase regimes with different center symmetry patterns. Phase transitions could also occur as we move along the parameter space of our theory, that is to say when altering $\beta$ and $am$. 
	
	Center symmetry breaking can be investigated by studying the corresponding order parameter. The  order  parameter  for  the  center  symmetry behaviour along the direction ${\hat \mu }$ is the vacuum expectation value  of the traced Polyakov loop in that direction: 
	\begin{eqnarray}
	P_{\mu} = \frac{1}{N}\sum_{x_{\perp}} \mathrm{Tr} \left(\prod_{i=0}^{L_{\mu}-1} U_{\mu}(x_{\perp},x_i)\right)\,,
	\end{eqnarray}
	where $x_i$ is the coordinate along the ${\mu}$-th direction, $x_{\perp}$
	the set of coordinates in the perpendicular directions to $\hat{\mu}$
	and $L_{\mu}$ the number of lattice points in the $\hat{\mu}$
	direction. {\color{black} The temporal direction is denoted by $\mu=0$ while the three spatial directions by $\mu=1,2,3$.}
	
	For large enough physical lattice volume, the center symmetry should be unbroken and the distribution of   $P_{\mu}$  along all directions should show  a single  peak  at  zero. This pattern will change as we reduce the lattice size. In a center symmetry broken phase, at smaller lattices, $N$ peaks will start to appear in one of the Polyakov loop distributions. However,  also more complicated breaking are possible for fermions in the adjoint representation depending on lattice volume and boundary conditions. Only in the case of thermal boundary conditions a single phase transition is expected to a deconfined phase with broken center symmetry.\footnote{Thermal boundary conditions refers to periodic boundary conditions for all fields, except for the fermions. These have antiperiodic boundary conditions in one direction with the smallest extend corresponding to the inverse temperature.} For periodic boundary conditions different intermediate phases can occur and a reconfined phase with unbroken center symmetry is observed if the lattice size in one direction becomes very small~\cite{Cossu:2009sq}. The intermediate phases might even be absent at small fermion masses as observed in Super Yang--Mills theory~\cite{Bergner:2014dua,Bergner:2018unx}. 
	
\begin{figure}[h]
	\includegraphics[width=7cm]{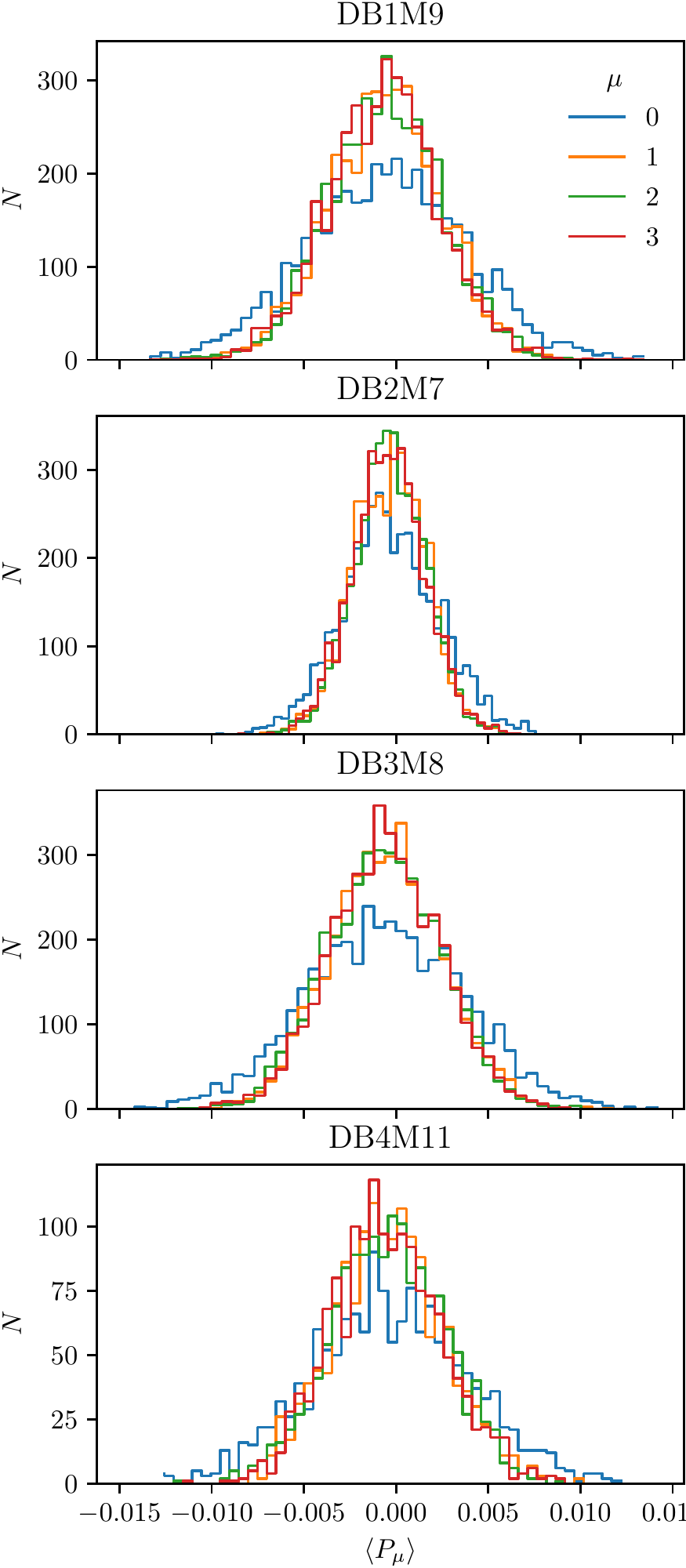}
	
	\caption{ The  histograms  of  the  average Polyakov loop for the ensembles DB1M10, DB2M7, DB3M8 and DB4M11 belonging to the set shown in  the  label  of  each  subfigure,  for  the four  space-time  directions. Single peaks indicate an unbroken center symmetry.}
	\label{fig:Polyakov_distribution}
\end{figure}

	We consider periodic boundary conditions in all spacial directions and the significantly larger temporal direction can be neglected in this consideration. None of the ensembles shows any indications for center symmetry breaking. To demonstrate this, in Figure~\ref{fig:Polyakov_distribution} we present histograms of the vacuum expectation value of Polyakov loops along all four directions for the ensembles DB1M10, DB2M7, DB3M8, and DB4M11. These  are the configurations which correspond to the four values of $\beta$, and the smallest mass for each. 
	
	In order to exclude any possibility that this signal might accidentally be due to a reconfined phase not connected to confinement at large volume, we have done some additional investigations. In such a phase center symmetry breaking is expected when the periodic boundary conditions are changed to thermal ones. We have confirmed, in additional simulations using thermal boundary conditions, that these small-volume effects are absent in our calculations.

\subsection{Topological Aspects}
\label{sec:topology}
The second important aspect that needs to be considered in order to check the validity of the considered parameter range are topological charge fluctuations of the ensembles. Our simulations should be ergodic and hence effectively explore all topological sectors. In Monte-Carlo simulations based on HMC algorithms strong autocorrelations of the topological charge appear as one approaches the continuum limit.  This well-known phenomenon of topological freezing has to be excluded for the extended $\beta - am$ parameter space that we are considering in our current simulations. 

In the continuum, the topological charge $Q$ is defined as the integral over the four-dimensional volume of the topological charge density
	\begin{eqnarray}
	Q = \frac{1}{32\pi^2} \int d^4 x \: \epsilon_{\mu\nu\rho\sigma} \Tr\left[G_{\mu\nu}(x)G_{\rho\sigma}(x)\right] \,.
	\label{eq:Q_continuum_def}
	\end{eqnarray}
	The discretized lattice counterpart of the above quantity can be obtained by replacing the gluonic field strength with a lattice operator that reproduces the correct continuum limit. The choice is not unique, and operators with better finite-size effects can be obtained by using $O(a)$-improved discretizations of the field strength. For the purposes of this work we incorporated the so-called ``clover'' definition,
	\begin{eqnarray}
	Q_L = \frac{1}{32\pi^2} \sum_{x} \epsilon_{\mu\nu\rho\sigma} \Tr\left[C_{\mu\nu}(x)C_{\rho\sigma}(x)\right]\,,
	\label{eq:Q_fieldteo_def}
	\end{eqnarray}
	where the ``clover leaf'' \(C_{\mu\nu}\) corresponds to the sum of the plaquettes \(P_{\mu\nu}(x)\) centered in \(x\) and with all the possible orientations in the \(\mu\nu\)-plane. This operator is even under parity transformations and exhibits \(\mathcal{O}(a^2)\) discretization effects. After a certain smoothing of the gauge configurations, $Q$ tends to integer values labelling the topological sector for each gauge configuration.
	
	Smoothing is required to suppress ultraviolet fluctuations of the gauge fields that would contaminate measurements of $Q$.  We use the gradient flow \cite{Luscher:2010iy} with the standard Wilson action to implement it.\footnote{The elementary integration step is \(\epsilon=0.01\). The flow time $t$ is chosen large enough to cancel discretization effects while still keeping the topological content of the gauge fields unchanged. For this reason we extracted the topological charge for flow times \(\tau_{\rm flow} =  t_{\rm flow} / a \) corresponding to $\sqrt{8t_{\rm flow}} = L/2$.}

\begin{figure}[h]
		\includegraphics[width=\columnwidth]{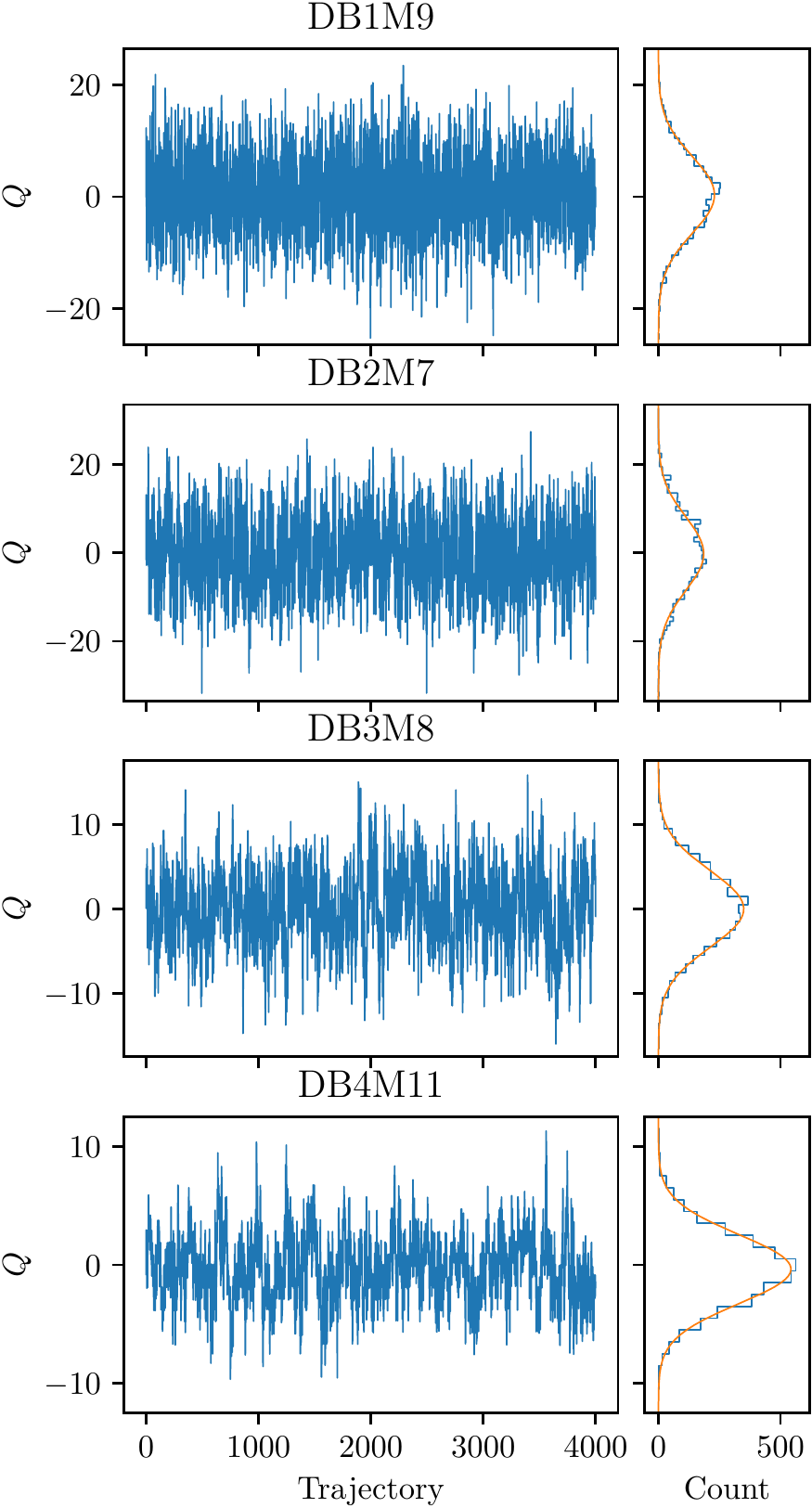}
		
		\caption{\color{black} Monte  Carlo  history  of $Q$ (left  panel),  and  resulting  histogram  (right  panel)  computed  on 4000 configurations extracted using gradient flow at gradient flow time $\tau$ corresponding to $a \sqrt{8 \tau_{\rm flow}} = L/2$ for the ensembles DB1M10, DB2M7, DB3M8 and DB4M11.}
		\label{fig:topology_histogram_distribution}
	\end{figure}

	To monitor possible effects of  topological freezing and, thus, a loss of ergodicity, we investigate the Monte-Carlo history and the histogram of the topological charge for several different parameters.
	In Figure \ref{fig:topology_histogram_distribution} we present the Monte-Carlo history of the topological charge $Q$ for the ensembles DB1M10, DB2M7, DB3M8, and DB4M11 with the corresponding histogram. These ensembles correspond to the smallest fermion masses at the four different values of $\beta$. One  can  observe  that  for none of the ensembles an extreme  autocorrelation  time  of $Q$ is observed and topological freezing is hence not a problem in the considered parameter range. A large number of topological sectors are explored and the resulting histogram is compatible with a Gaussian distribution with average at zero. This is a strong indication for a adequate sampling of topological observables. Full results for the autocorrelation time, and for the fitted shape of the histogram of $Q$, are shown in Appendix~\ref{sec:app_results}.

    Besides the check for the reliablility of the simulations, this study provides also further insights into the physical properties of the theory.
    The topological susceptibility $\chi$,
    \begin{eqnarray}
        \chi = \frac{\langle Q^2 \rangle  -  \langle Q \rangle^2 }{V} \,,
    \end{eqnarray}
    provides information about the possible IR conformal scenario for the theory.

    An IR conformal theory becomes confining when deformed by a fermion mass. If this occurs, the theory should be indistinguishable from the equivalent quenched theory. A good quantity to test this scenario is the topological susceptibility, since it should match the pure \su{2} Yang--Mills  theory~\cite{Bennett:2012ch} results. 
    
\begin{figure}
		\includegraphics[width=\columnwidth]{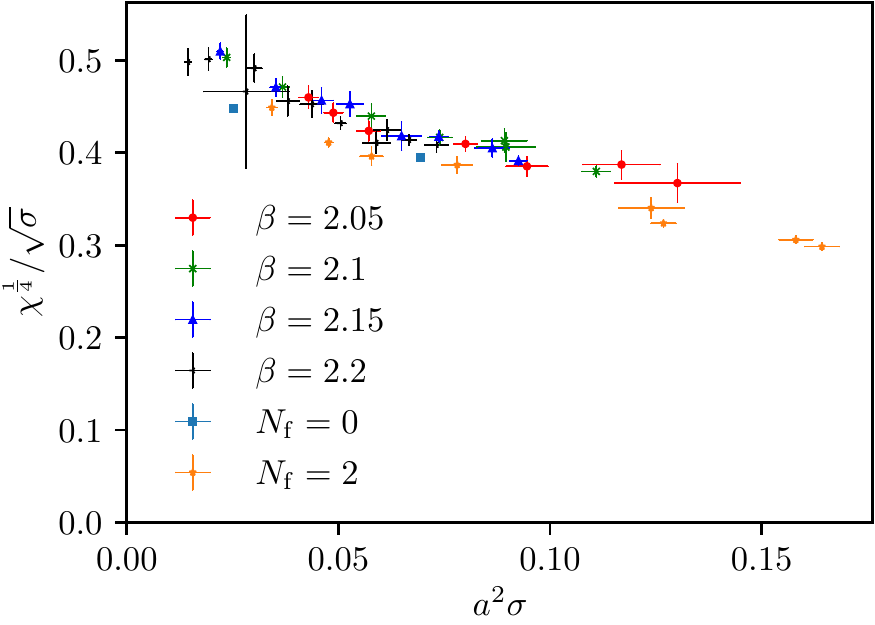}
		
		\caption{The topological susceptibility measured for the largest combination of each ($\beta$, $m$), in units of the string tension $\sqrt{\sigma}$. Included for comparison are data for pure $\su{2}$
		gauge theory and the $\su{2}$ theory with two adjoint Dirac flavors,
		using data from \cite{Bennett:2012ch}.}
		\label{fig:susceptibility}
	\end{figure}

    In Fig.~\ref{fig:susceptibility} we present the topological susceptibility in units of the string tension $\chi^{\frac{1}{4}}/  \sqrt{\sigma}$ as a function of the string tension in lattice units $a^2 \sigma$. This is compared to the same quantity in pure \su{2} Yang--Mills theory corresponding to the quenched $N_f=0$ case. Furthermore the result for the $N_f=2$ case are shown, which is expected to be IR conformal. The topological susceptibility is in good agreement between these three theories; {\color{black} any discrepancies are understood as higher order scaling corrections of $a^2 \sigma$.}   This provides a first indication that \OFAQCD{} is IR conformal.
	
	The gradient flow also enables the definition of scale parameters $t_0$ and $w_0$, which can be determined quite accurately to high precision.  These flow observables have been introduced  in~\cite{Luscher:2009eq,Borsanyi:2012zs}. $t_0$ is defined according to the following prescription. First, we set
    \begin{eqnarray}
        F(t) = t^2 \langle E(t) \rangle \, \ {\rm where} \ E(t) = \frac{1}{4} B^2_{\mu \nu} (t)\,,
    \end{eqnarray}
    where $B_{\mu \nu}$ is field strength obtained by flowing $G_{\mu \nu}$ along the flow time direction. We identify the scale $t_0$ as the value of $t$ for which
    \begin{eqnarray}
        F(t) |_{t=t_0(c)} = c\,. 
    \end{eqnarray}
    Similarly $w_0$ is defined by
     \begin{eqnarray}
        t \frac{\mathrm{d} }{\mathrm{d}t}F(t) |_{t=w^2_0(c)} = c\,. 
    \end{eqnarray}
    In these definitions $c$ is chosen such that the relevant condition $a \ll \sqrt{8 t_0} \ll L$ or $a \ll \sqrt{8} w_0 \ll L$ is satisfied. Small $c$ lead to larger lattice artefacts and larger $c$ usually lead to larger autocorrelations \cite{Bergner:2014ska}. In our case we take $c=0.2$ consistent with the common value of $c=0.3$ for QCD assuming a scaling with $N$ as in \cite{Ayyar:2017qdf}. $G_{\mu \nu}$ is represented by the clover plaquette.
	
\subsection{Particle spectrum}
\label{sec:particle}
The first important observables which we consider to determine conformal or confining infrared behaviour are the masses of the particles. We considered masses of mesonic and baryonic two-fermion states, glueballs, and a \spinhalf {\color{black}  hybrid fermion} state. 
In addition, we have determined string tension and scale setting quantities ($w_0$ and $t_0$).
In this section we first discuss basic features of these data which are later on considered in more detailed fits according to a conformal or confining scenario. In the current work, we focused on the most precise data and excluded two-fermion states with disconnected contributions. Our methodology for determining the masses follows very closely our previous work~\cite{Athenodorou:2014eua}. 
\begin{figure}
  \includegraphics[width=\columnwidth]{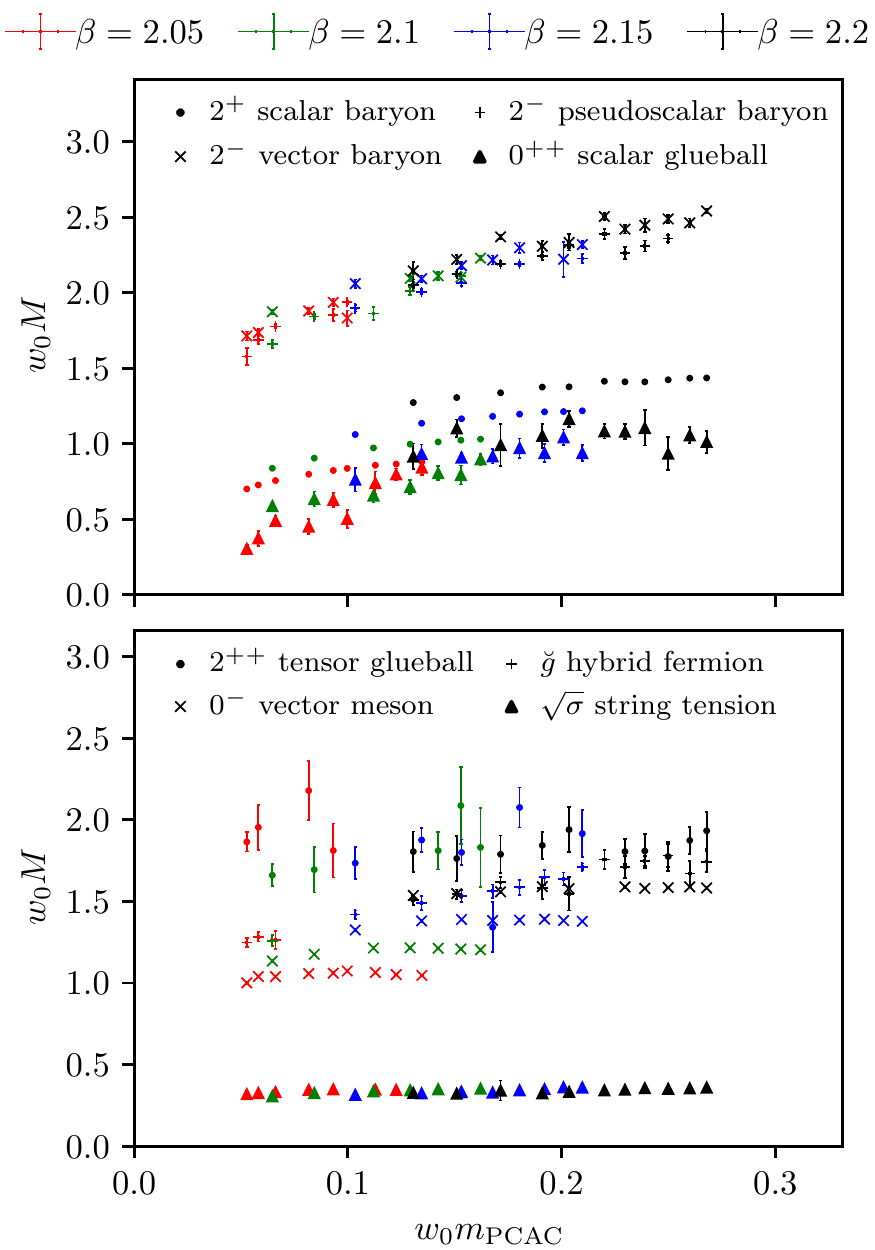}

  \caption{The mass spectrum observed at the largest volume for each combination of ($\beta$, $m$), showing baryons, glueballs, the hybrid fermion $\breve{g}$, and the string tension $\sqrt{\sigma}$, as a function of the PCAC fermion mass, in units of the gradient flow scale $w_0$.}
  \label{fig:massspectrum}
\end{figure}
\begin{figure}
  \includegraphics[width=\columnwidth]{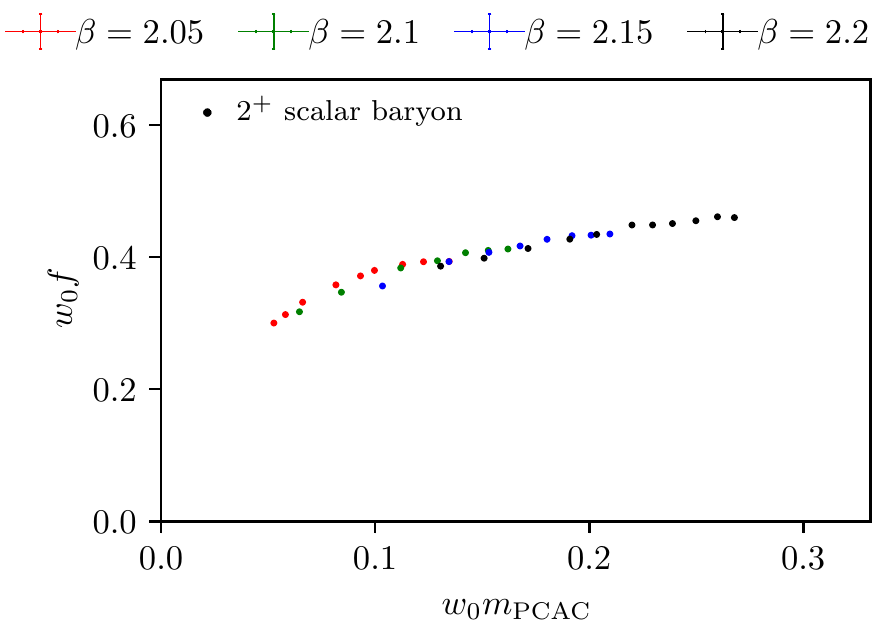}

  \caption{The decay constant of the $2^+$ baryon observed at the largest volume for each combination of ($\beta$, $m$), as a function of the PCAC fermion mass, in units of the gradient flow scale $w_0$.}
  \label{fig:decayconst}
\end{figure}

The data are summarized in Appendix~\ref{sec:app_results}  and represented in units of $w_0$ in Figures \ref{fig:massspectrum} and \ref{fig:decayconst}. In order to control finite volume effects, we have included different volumes for some parameters.

Several important observations can already be made from this representation of the data. As observed in our previous study, there is only a small dependence of the masses in physical units on the PCAC mass {\color{black} which becomes more profound as $\beta$ and PCAC mass increase.} This is consistent with the expectation for a conformal theory. In such a case, masses of states and quantities used for scale setting tend to zero in the chiral limit, {\color{black} i.e as $m_{\rm PCAC} \to 0$}, with the same exponent, leading to constant mass ratios and constant masses in units of $w_0$. 

At small PCAC masses, there is still a considerable deviation from the expected constant behaviour. In particular the scalar baryon ($\gamma_5$ channel), which should become Goldstone boson in case of chiral symmetry breaking, tends towards smaller masses. However, this trend is reduced at larger values of $\beta$ and might hence be related to lattice artefacts. More detailed considerations are therefore required to resolve the continuum behaviour of the theory.

\begin{figure}
    \centering
    \includegraphics{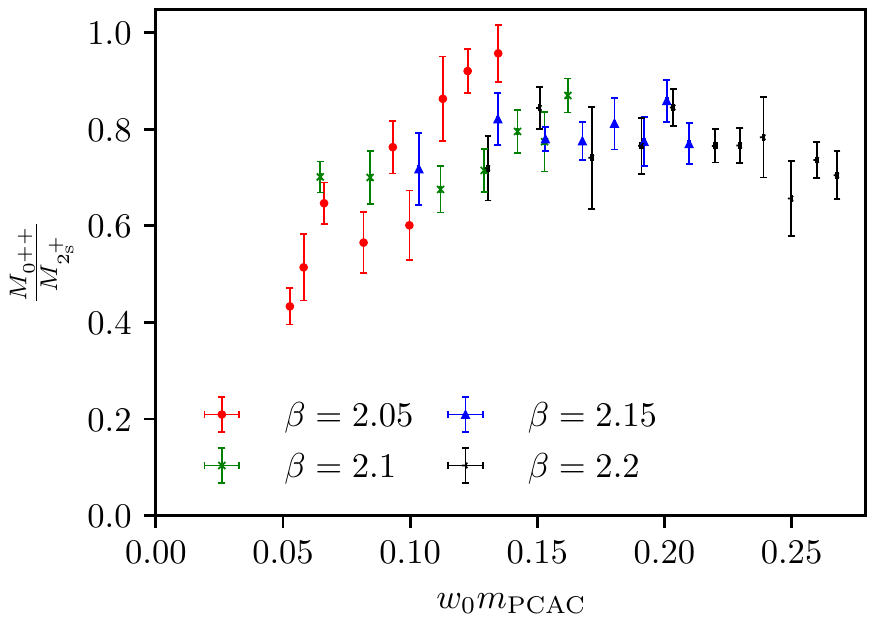}
    \caption{The ratio of the mass of the $0^{++}$ glueball to that of the $2^+$ scalar baryon as a function of the PCAC fermion mass scaled by the gradient flow scale $w_0$.}
    \label{fig:scalar-ratio}
\end{figure}

The ordering of the states in the particle spectrum provides further insights in possible infrared scenarios for the theory. Different from a chiral symmetry breaking scenario, the scalar glueball is the lightest state of the theory, lighter than the $2^+$ scalar baryon at all points observed, as shown in Figure~\ref{fig:scalar-ratio}. A light glueball is also observed in several near-conformal theories, and might even be related to a dilaton state in the conformal limit. The \spinhalf {\color{black} {hybrid fermion}} state is rather light compared to other baryonic and mesonic states. However, it is still heavier than the scalar glueball and scalar baryon. Therefore our results rather indicate that the discussed exotic infrared scenario conjectured in \cite{Anber:2018tcj,Bi:2018xvr} with light baryon states is not realized for this theory.

\subsection{Determination of the mass anomalous dimension for the different gauge couplings}
\label{sec:anomalous}
We continue the analysis considering an IR conformal scenario of \OFAQCD{}. In such a scenario, the scaling of relevant observables is governed by the mass anomalous dimension at the IR fixed point $\gamma_*$. 
In the following, we consider two methods for the determination of the scaling. The first one is a hyperscaling analysis of the particle spectrum. The second one considers the scaling of the mode number, the integrated eigenvalue density of the Dirac operator. Consistent fits are a good indication for the realization of a conformal scenario. In this section we first analyse separately the different values of the gauge coupling. A final result of this quantity at the fixed point would be independent of the gauge coupling. In the last part we consider possible ways to determine the universal scaling based on the data of the complete parameter range.

In our previous investigation we have determined the value at a single $\beta=2.05$~\cite{Athenodorou:2014eua}. This value is updated with additional ensembles at the same gauge coupling.

\subsubsection{Particle spectrum}
Near-conformal behaviour would indicate a particular scaling of the particle masses. Including the approximate finite-size effects, a spectral quantity $M$ as a function of the $\mpcac$ mass, 
follows the scaling relation~\cite{DeGrand:2009mt,DeGrand:2009hu,Lucini:2009an,DelDebbio:2010hx,DelDebbio:2010hu,DelDebbio:2010ze,DelDebbio:2010jy}  
\begin{equation}
	L a M = f \left(L\left(a\mpcac\right)^{\frac{1}{1+\gamma_*}} \right)\;,
	\label{eq:functiof}
\end{equation}
for some function $f$, where $L\rightarrow \infty$ is the finite
spatial extent of the lattice. This implies that the points in a plot of
$LaM$ against $L\left(a\mpcac\right)^{1/(1+\gamma_*)}$ collapse to one universal curve
for the correct value of $\gamma_*$. 

\begin{figure}
  \includegraphics[width=\columnwidth]{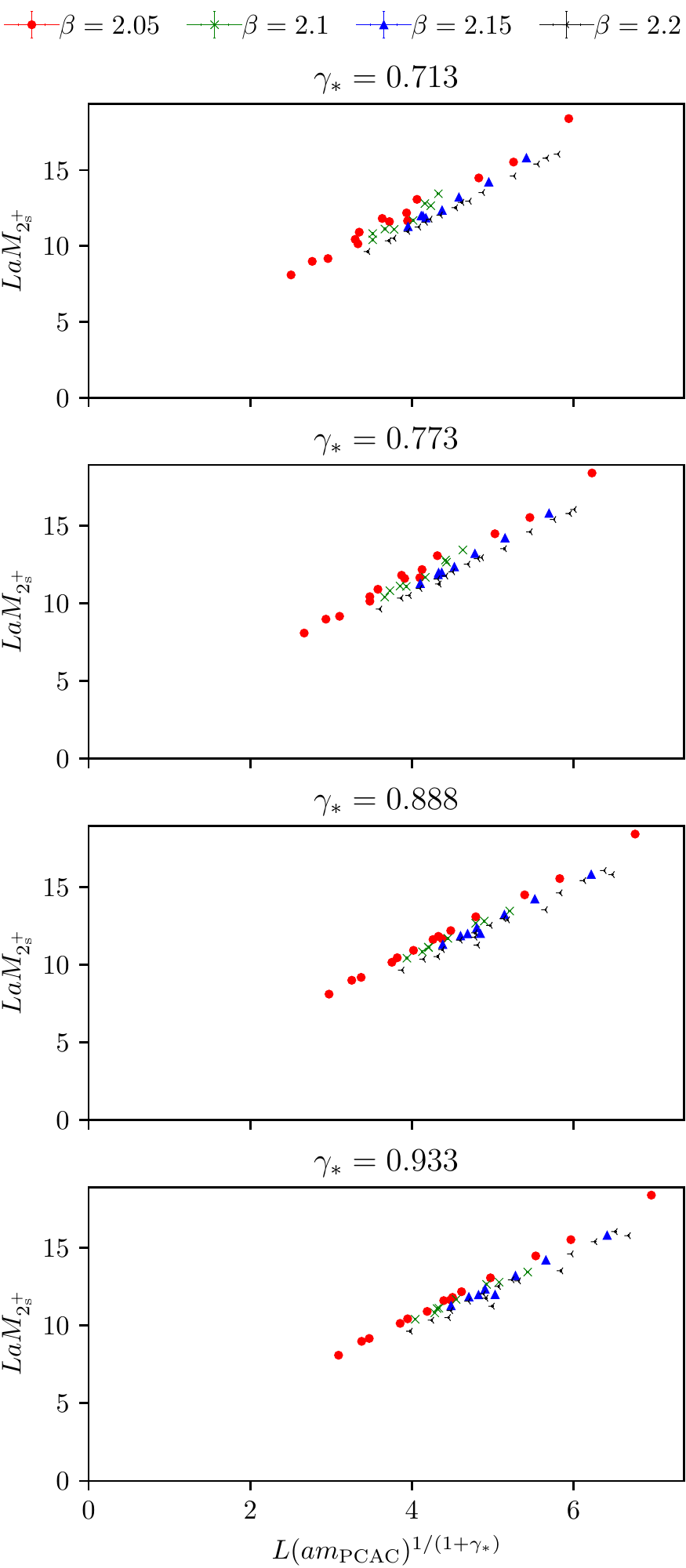}

  \caption{Finite-size hyperscaling plots showing the mass of the $2^+$ baryon as a function of the product $L\mpcac^{1/(1+\gamma_*)}$ for the stated values of $\gamma_*$. Data for different values of $\beta$ align at different values of $\gamma_*$, with the extrema being $\beta=2.05$ data aligning at $\gamma_*=0.943(71)$, and $\beta=2.2$ aligning at $\gamma_*=0.717(43)$.}
  \label{fig:fshs}
\end{figure}

These plots, based on the data for the scalar baryon, are shown in Figure~\ref{fig:fshs}. In a certain range of $\gamma_*$ the values for all gauge couplings become nearly consistent with a universal function. 
In order to perform a more precise determination, we have applied the methodology proposed by DeGrand \cite{DeGrand:2009hu} based on work by Bhattacharjee and Seno \cite{Bhattacharjee_2001}.
The principle of this method is to allow the data to shape the fit function, rather than imposing a fitting form. This is done by minimisation of the function
\begin{equation}
    P(\gamma_*) = \frac{1}{N_\mathrm{o}} \sum_{\ell \in \{L\}} \sum_{i\in S_{\ell}} \left(L_i aM_i - f_{\ell}\left(L_i^{1+\gamma_*} am_{\mathrm{PCAC}\,i} \right)\right)^2\;,
\end{equation}
where $S_{\ell}$ is the set of data such that for $i\in S_{\ell}$, $L_i \ne \ell$, and $L_i^{1+\gamma_*} am_{\mathrm{PCAC}\,i}$ is in the range spanned by the data for which $L=\ell$; and $N_\mathrm{o}$ is the number of points $i$ considered in total, i.e.~$N_\mathrm{o} = \sum_{\ell\in\{L\}} \sum_{i\in S_{\mathrm{\ell}}} 1$. $f_{\ell}$ is a function interpolating the values of $LaM$ for data for which $L=\ell$; we choose this interpolation to be piecewise linear. This minimisation pulls the data for each value of $L$ to the others, collapsing the data to a single global function without needing to specify its functional form.

This detailed analysis reveals more significant deviations between $\gamma_*$ obtained at different values of the gauge coupling $\beta$. As shown in Table~\ref{tab:fshs-gamma} smaller values of $\gamma_*$ are preferred at larger $\beta$.

\begin{table}[h]
    \centering
    \begin{tabular}{c|cc}
    $\beta$ & $\gamma_*$ & $N_{\mathrm{points}}$ \\
    \hline
    \hline
    2.05 & $0.933(71)$ & 14 \\
    2.1 & $0.89(12)$ & 4 \\
    2.15 & $0.773(47)$ & 7 \\
    2.2 & $0.713(43)$ & 20
\end{tabular}

    \caption{Results of fitting the anomalous dimension $\gamma_*$ from finite-size hyperscaling using the methodology described in Refs.~\cite{DeGrand:2009hu,Bhattacharjee_2001}. $N_{\mathrm{points}}$ is the total number of points contributing to the residual at the given value for $\gamma_*$.}
    \label{tab:fshs-gamma}
\end{table}

\subsubsection{Mode number}
The eigenvalue spectrum of the Dirac operator allows a more precise determination of $\gamma_*$ compared to the scaling of the particle spectrum. We have already described the method in~\cite{Athenodorou:2014eua}.
The integrated eigenvalue density $\nubar(\Omega)$ or mode number as a function of upper integration limit
$\Omega$ follows in an intermediate regime the scaling
\begin{equation}
	a^{-4}\nubar(\Omega) \approx a^{-4} \nubar_0(\tilde{m}) + A[(a\Omega)^2-(a\tilde{m})^2]^{\frac{2}{1+\gamma_*}}\;,
\label{eq:nubar-scaling}
\end{equation}
where $\tilde{m}$ is some renormalized fermion mass  \cite{Patella:2012da,Cheng:2013eu}. 

\begin{figure}
  \includegraphics{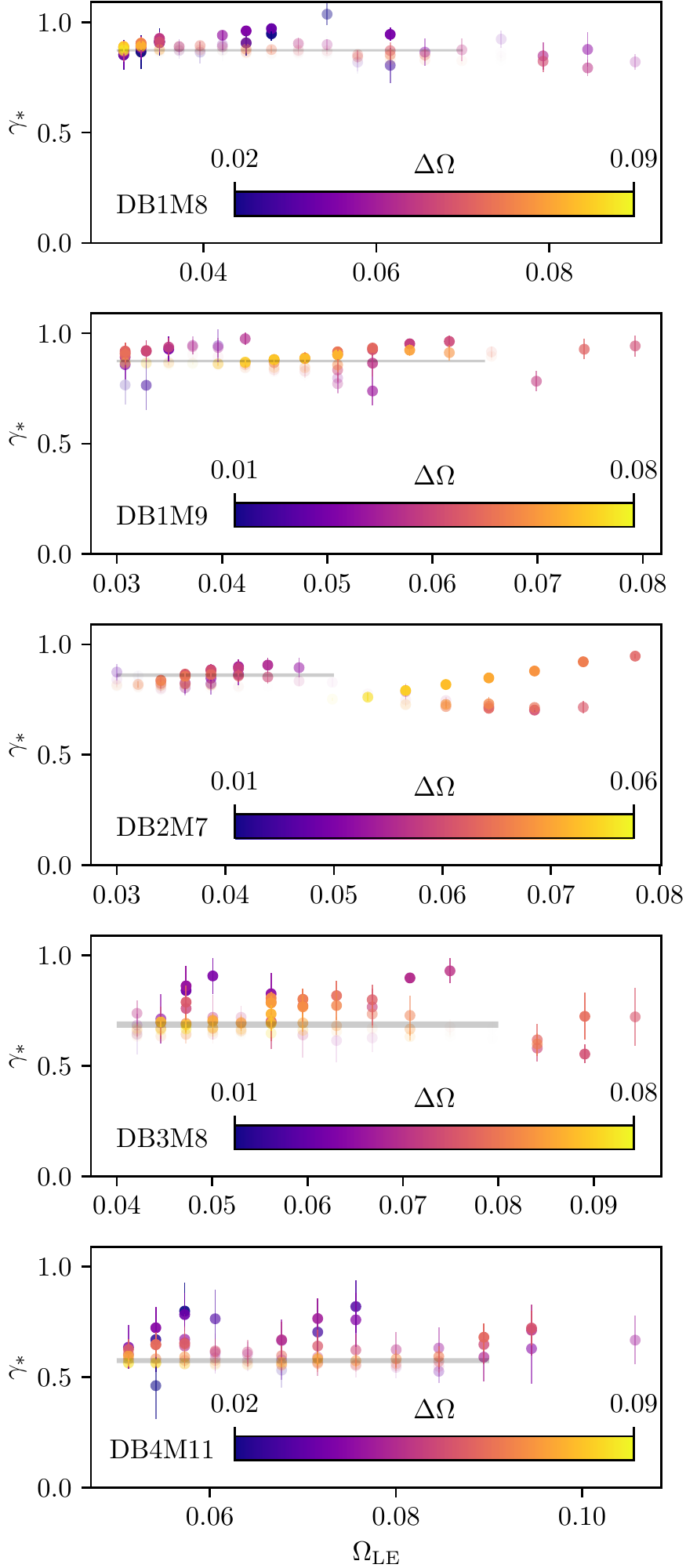}
  
  \caption{Results of the bootstrapped fits of the mode number for the ensembles DB1M8, DB1M10, DB2M7, DB3M8, and DB4M11 as a function of the lower end of the fit window. Colors represent the length of the window $\Delta\Omega = \Omega_{\mathrm{UE}} - \Omega_{\mathrm{LE}}$, and opacity represents the proportion of fits for that window that converged.}
  \label{fig:modenumber}
\end{figure}

To find the region of validity, we have used several different windows $[\Omega_{\mathrm{LE}},
\Omega_{\mathrm{UE}}]$ for the fit according to the scaling formula \eqref{eq:nubar-scaling}.
The numerical fit and the specific functional form leads to an additional dependence of the results on the initial parameters. In order to
obtain an estimate of this uncertainty, many repeated fits are performed with small variations in the initial parameters around a central value.
This central value is chosen by a constrained fit in which
$\nubar_0=0$, which is significantly more stable than the full four-parameter fit.
Examples of the scattering of $\gamma_*$ obtained in these fits are shown in Figure \ref{fig:modenumber}.

We have determined optimal values for $\gamma_*$ by searching for a plateau in the distribution, constraining both $\Omega_{\mathrm{LE}}$ and $\Delta\Omega=\Omega_{\mathrm{UE}}-\Omega_{\mathrm{LE}}$, and for the most reliable fits.
The plateau is then fitted, weighting the result for each window both by the statistical uncertainty and also by the number
of fits of that window which converged.
The determination is done separately at each $\beta$ focusing in particular at the large volumes and small fermion masses.
The final results are shown in Table~\ref{tab:gammamode}.
\begin{table}
    \centering
    \begin{tabular}{c|cccc|c}
    Ensemble & $\Omega_{\mathrm{LE}}^{\mathrm{min}}$ & $\Omega_{\mathrm{LE}}^{\mathrm{max}}$ & $\Delta \Omega_{\mathrm{min}}$ & $\Delta \Omega_{\mathrm{max}}$ & $\gamma_*$ \\
    \hline
    \hline
    DB1M8 & 0.03 & 0.07 & 0.05 & 0.1 & $0.8720(52)$ \\
    DB1M9 & 0.03 & 0.065 & 0.05 & 0.1 & $0.8743(56)$ \\
    DB2M7 & 0.03 & 0.05 & 0.02 & 0.08 & $0.8609(83)$ \\
    DB3M8 & 0.04 & 0.08 & 0.05 & 0.08 & $0.687(14)$ \\
    DB4M11 & 0.05 & 0.09 & 0.05 & 0.09 & $0.575(11)$
\end{tabular}

    \caption{The values of the mass anomalous dimension $\gamma_*$ obtained from a fit of the mode number data according to \eqref{eq:nubar-scaling}.}
    \label{tab:gammamode}
\end{table}

\subsubsection{Universal behavior or dependency on the gauge coupling?}
The obtained values of the anomalous dimension using the expected scaling of the particle spectrum or the mode number are quite consistent. The deviations of the two methods are between 0.2 and 3 sigma with the smallest deviation at $\beta=2.1$ and the largest at $\beta=2.2$. The large deviation at  $\beta=2.2$ might be seen as an indication that our determination of the scaling is not completely under control in this case. However, deviations at the order of 3 sigma between these two methods have also been observed in studies with two Dirac flavors.

An even more striking result is the dependency of $\gamma_*$ on the value of $\beta$: for nearly-conformal gauge theories, the fixed point value of the anomalous dimension is a universal quantity, and the gauge coupling is at most marginal. In our calculation instead we observe that the larger $\beta$
indicate a smaller value of $\gamma_*$. This may indicate that the theory is not IR conformal, despite the observed IR scaling seem compatible at first sight with that hypothesis. However, large scaling corrections or strong lattice artefacts due to the nearby bulk transition might provide an alternative explanation. We now discuss these two possibilities. 

Assuming the deviation can be explained by scaling corrections, a fit approach of \cite{Cheng:2013xha} can be applied to quantify leading and sub-leading contributions. We have tried to fit our data according to this approach. However, we did not obtain consistent result since the sign of the leading exponent did not agree with the assumptions for an IR fixed point.

This leaves us with one basic interpretation of the data, which is a rather drastic correction due to the lattice artefacts. Since we have chosen our values of the gauge coupling as close to the bulk transition as possible, such effects are not unexpected. The conclusion of this line of argument is that the largest $\beta$ provides the most reliable result of the conformal scaling and $\gamma_*$ is smaller than estimated in our earlier work.

\subsection{Chiral perturbation theory}
\label{sec:chipt}
In order to check the infrared scenario for the theory, the assumptions of a near-conformal scaling must be compared with the chiral symmetry breaking case. Assuming chiral symmetry breaking, the functional dependence is provided by chiral perturbation theory. We have performed a basic fit of the scalar baryon state according to this scenario. This state corresponds to the pseudo-Nambu-Goldstone boson of broken chiral symmetry. 

\begin{figure}
  \includegraphics[width=\columnwidth]{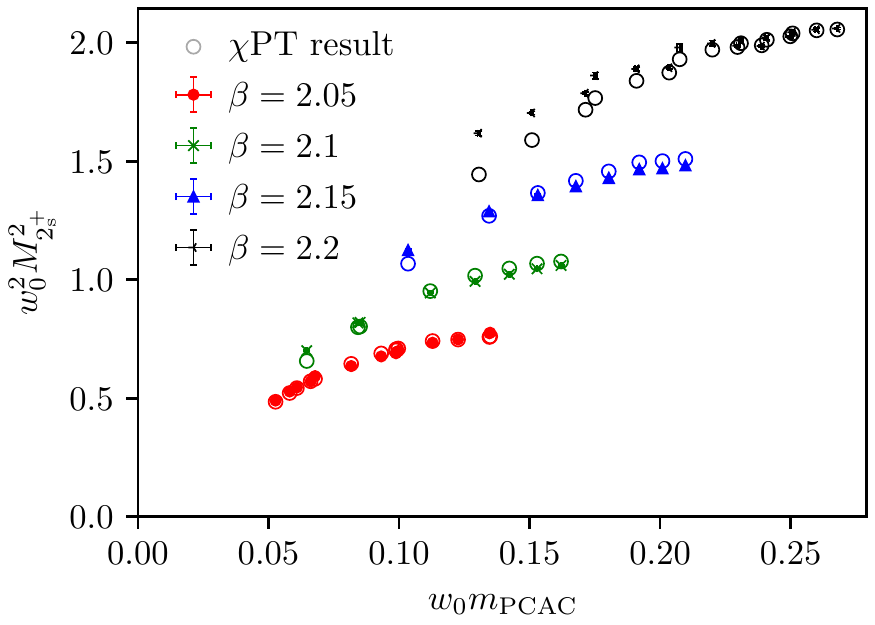}

  \caption{The results of fitting the $2^+$ baryon mass (scaled by the gradient flow scale $w_0$) for the largest volume ensemble for each combination of ($\beta$, $m$) with chiral perturbation theory.}
  \label{fig:Xpt}
\end{figure}

The fit has been done according to the functional form 
\begin{align}
    \hat{M} =& 2B\hat{m}_{\mathrm{PCAC}}(1+L\hat{m}_{\mathrm{PCAC}}+D_1 \hat{m}_{\mathrm{PCAC}} \log(D_2 \hat{m}_{\mathrm{PCAC}})) \nonumber \\
    &+ W_1 a\mpcac \nonumber \\
    &+ W_2 \frac{a^2}{w_0^2}\,,
\end{align}
where $\hat{m}_{\mathrm{PCAC}} = w_0\mpcac$, fitting the coefficients $B$, $L$, $D_1$, $D_2$, $W_1$, and $W_2$. The result is presented in Fig.~\ref{fig:Xpt}.
It indicates rather large lattice artefacts and the strongest deviation from the fit is seen at the largest $\beta$. Assuming the standard confinement and chiral symmetry breaking scenario, the largest $\beta$ corresponds to the smallest lattice spacing.

To exclude the possibility that this fit has been anchored by the coarsest lattice spacing, where the scalar baryon mass does not exhibit the near-flat behavior observed at larger values of $\beta$, we have performed additional fits of only the finer values of the lattice spacing. This fit similarly shows very similar lattice artifacts to the fit including data at all values of $\beta$, suggesting that the inconsistency with chiral perturbation theory is a genuine effect, and not an artefact of the choice of fit region.

Further simulations with larger volumes and smaller masses at this $\beta$ might help to reduce the uncertainty.
Overall chiral symmetry breaking cannot be excluded from these fits. However, the complete functional dependence of the data is not well described by chiral perturbation theory. In addition the fact that the scalar glueball  is lighter than the scalar baryon is also in contradiction with chiral symmetry breaking.

\subsection{$R$ ratio}
\label{sec:R_ratio}
In this section we consider our new data for \OFAQCD{} to investigate universal properties that can be used to characterize the general parameter space of strongly interacting gauge theories. In an earlier work~\cite{Athenodorou:2016ndx}, we have shown that the observable
\begin{eqnarray}
    R = \frac{M_{2^{++}}}{M_{0^{++}}} \, ,
\end{eqnarray}
where $M_{2^{++}}$ is the mass of the lightest spin-2 composite state and $M_{0^{++}}$ the mass of the lightest spin-0 state, is related to the anomalous dimension $\gamma_*$. It can be used to characterize general properties of a IR conformal gauge theory. Some interesting properties makes $R$ a particularly relevant observable. More in detail, $R$ is defined universally, can be computed explicitly for a wide range of models, is scheme-independent, and it is not directly controlled by internal global symmetries of the theory. Therefore, it is legitimate to compare the value of $R$ computed in theories with completely different internal symmetries and symmetry-breaking patterns. This is a particularly welcome feature in the context of gauge theories with fermionic field content, where the physics of chiral symmetry and its breaking introduces non-trivial model-dependent features.

Before presenting numerical results for $R$, we briefly review the expected behaviour for a conformal theory as it is deformed by some fermion mass $m$. For a region of small, finite deforming mass, spectral masses will scale as $M \propto m^{1/\Delta}$ where $\Delta=1 + \gamma_{\star}$ is the scaling dimension. A finite lattice volume introduces an absolute IR cutoff, thus,  an appropriate scaling variable to consider is $x = L m^{1/\Delta}$ and so the $i^{\rm th}$ mass scale would become a function of $x$, i.e.~$LM_i = f_i (x)$ as we demonstrated in Eq.~\eqref{eq:functiof}. To first order for a state of mass $M_0$, this gives $LM_0 \propto x$ and by back substitution we obtain $L M_i = f_i (LM_0) $. Thus if we take mass ratios, then the dependence on the lattice extent drops out, allowing us to compare data at different volumes
\begin{eqnarray}
\frac{M_i}{M_j} = \frac{f_i(LM_0)}{f_j(LM_0)}\,.
\end{eqnarray}
We expect that the behaviour of $R$ will be characterized by four distinct regimes. At large values of $m$, $R$ becomes effectively consistent with the value seen in the pure \su{2} Yang-Mills theory; this value has been found in lattice studies to be $R=1.44(4)$~\cite{Lucini:2001ej,Lucini:2004my}. At small $m$, the regime will depend on the lattice extent $L$. At large enough volumes the observed physics shows confining properties due to the mass deformation. At very small volumes, the theory observed is that of the so-called “femto-universe”~\cite{Athenodorou:2016ndx}; previous studies indicate that $R \simeq 1$ in this region~\cite{GonzalezArroyo:1988dz,Daniel:1989kj,Daniel:1990iz}. However, there is an intermediary region of $L$ where it is sufficiently large to observe the conformal behaviour, without being so large as to be distorted by confining effects. This is the region of interest, which can be extrapolated to a chiral limit value, allowing $R$ to be determined for the conformal case. Since $R$ enables comparison between disparate theories we choose to perform a comparison with a toy model constructed making use of the principles of gauge-gravity duality~\cite{Athenodorou:2016ndx}. This model is constructed within the bottom-up approach to holography, in such a way that the only operator deforming the theory and driving it away from conformality has scaling dimension $\Delta$. Given our results for the anomalous dimension, we can compare the measured ratio to expectations coming from the toy model with the corresponding value of $\Delta$. More details behind this idea can be found in our previous work~\cite{Athenodorou:2016ndx}.

\begin{figure}
  \includegraphics[width=\columnwidth]{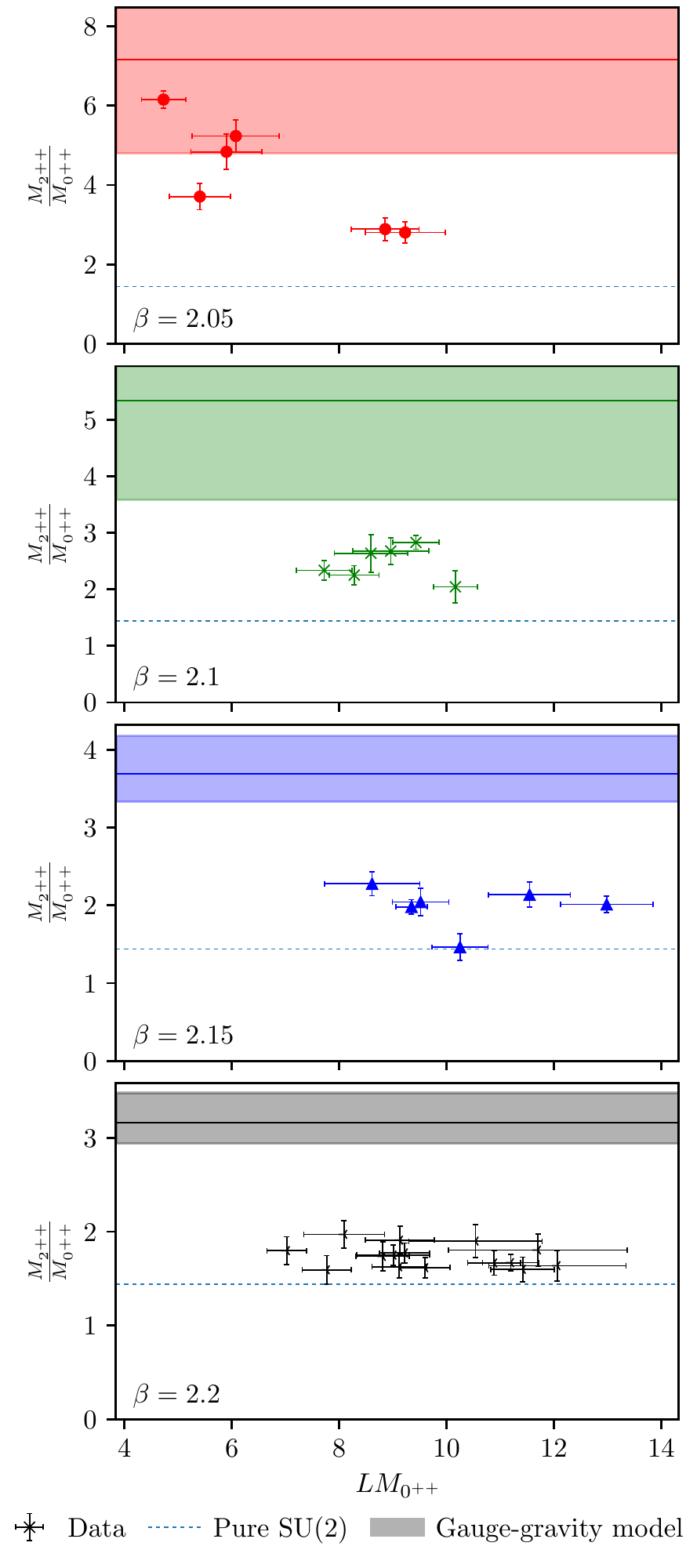}

  \caption{The ratio $R=\frac{M_{2^{++}}}{M_{0^{++}}}$ measured at all volumes studied for each combination of ($\beta$, $m$). The shaded region indicates the range suggested by the model presented in \cite{Athenodorou:2016ndx}, when the values for $\gamma_*$ in Tab.~\ref{tab:fshs-gamma} are used. In the case of $\beta=2.05$ and $2.1$ the upper bound on $R$ is infinite, as $\gamma_*$ is compatible with 1, at which point the model is singular.}
  \label{fig:Rratio}
\end{figure}

In Figure~\ref{fig:Rratio} we present our results on $R$ for the four different values of the coupling $\beta$. Our data are compared with the Ratio $R$ for the \su{2} pure Yang-Mills theory (blue dashed line) as well as with the prediction from the conformal string-inspired toy model (upper band). We remark that for each different value of $\beta$ the conformal prediction changes due to different measured values of $\gamma_*$. For large $LM_0$ the data for the four different $\beta$s moves towards to the pure gauge prediction, in agreement with the discussion above. For the case of our smallest value of $\beta$,  going to smaller $LM_0$, the calculations are becoming consistent with the conformal plateau with large $R$. For $\beta = 2.1$ and (to a more limited extent) for $\beta = 2.15$, there is some evidence of a behaviour that is different from the confining one, although the ratio does not raise enough to meet the value predicted by the model. At our largest value of $\beta = 2.2$, results do not move significantly above the Yang-Mills value. Further simulations are needed to confirm whether this tendency towards the confining behaviour is a reflection of the physical properties of the system or the data are not in the regime that can capture IR conformality. Turning now to the regime of very small volumes, and thus smaller values of $LM_0$, we observe that there are no signals of ``femto-universe". The above behaviour is in agreement with our previous findings in~\cite{Athenodorou:2016ndx}.

Once again, we see hints of different infrared behaviour at the lower and at the higher end of the $\beta$ range we have studied. In order to resolve the discrepancy, further simulations would be needed. In particular, for $R$ one would need to carefully design calculations that interpolate between the Yang-Mills behaviour at large $LM_0$ and the femto-universe at lower values of this quantity, with the goal of identifying an intermediate region where infrared conformality could be visible. 

\section{Conclusions}
\label{sec:con}
The $\su{2}$ gauge theory with one Dirac flavor in the adjoint representation is a very interesting theory due to its relation to supersymmetric counterparts and composite Higgs models. More interestingly, there are several complementary possibilities for the corresponding effective theory in the infrared limit. It could be similar to the case with two adjoint Dirac flavors, which shows significant indications for a conformal fixed point in the infrared. It could also follow a QCD-like scenario with chiral symmetry breaking in the infrared similar to supersymmetric Yang-Mills theory. As a third possibility a different infrared limit has been recently suggested based on 't Hooft anomaly matching. In general a small or even vanishing beta function is expected, which makes it rather difficult to disentangle the scenarios. From the theoretical point of view the fact that the theory is borderline between a conformal or non-conformal case, makes it an even more interesting subject. On the other hand, it provides severe challenges for the investigation on the lattice.

We have studied the theory at different lattice sizes and different values of the gauge coupling. We have investigated a large number of observables in order to characterize the infrared properties of the theory. These include the particle spectrum, topology, the Polyakov loops, and the mode number of the Dirac operator. We have confirmed the general reliability of the simulations since the gauge coupling $g$ is chosen below the bulk transition and there is no indication for deconfinement or topological freezing. For several observables we observe a significant dependence on the gauge coupling, which is unexpected for an infrared conformal scenario. There are basically three possible explanations.

\begin{enumerate}
    \item The theory is not conformal but rather QCD-like, in the infrared. Chiral symmetry is broken by a non-vanishing adjoint fermion condensate. For this conclusion, the larger deviations in the chiral fits need to be explained by lattice artefacts. It is in contradiction with the observed ordering of the particle spectrum since one would expect the Goldstone modes to be the lightest states. This could, however, be explained with too heavy fermion masses in the current simulations.
    \item The theory is conformal with scaling deviations. We are, however, currently unable to determine these violations in a reliable fit. It is in this case not completely clear which value of the gauge coupling is most reliable and brings us closest to a conformal fixed point.
    \item The theory is conformal, but the influence of the bulk phase is stronger than expected. In this case one would prefer the largest $\beta$ and, consequently, a smaller mass anomalous dimension is obtained.
\end{enumerate}
While not fully conclusive, our calculations {\color{black} based primarily on the ordering of the scalar baryon vs the scalar glueball (Fig~\ref{fig:scalar-ratio})} seem to disfavour scenario (1), therefore hinting towards the IR (near-)conformality of the theory. However, given the dependence of the mass anomalous dimension on the coupling, further investigations at large $\beta$ and at smaller fermion masses are needed in order to fully answer the question of the phase realised by the theory. 

Concerning the suggested alternative scenario based on 't Hooft anomaly matching, we can provide some indications that it might not be realized in the infrared. In this case, one would expect that the \spinhalf {\color{black} hybrid} fermions become the lightest states at small fermion masses. Instead we observe that these states are of the same order as the vector meson mass and heavier than the scalar bayon and the scalar glueball.

\begin{acknowledgments}
    We thank M.~Piai and D.~Elander for providing the data on the $R=M_{2^{++}}/M_{0^{++}}$ for the string inspired toy model. We also thank M.~\"Unsal for discussions.  
   AA has been financially supported by the European Union's Horizon 2020 research and innovation programme ``Tips in SCQFT'' under the Marie Sk\l odowska-Curie grant agreement No.~791122. The work of EB has been funded by the Supercomputing Wales project, which is part-funded by the European Regional Development Fund (ERDF) via Welsh Government, and by the Japan Society for the Promotion of Science under award PE13578.
   The work of BL has been supported in part by the European Research Council (ERC) under the European Union’s Horizon 2020 research and innovation programme under grant agreement No 813942 and by the UKRI Science and Technology Facilities Council (STFC) Consolidated Grant ST/P00055X/1. The work of BL is further supported in part by the Royal Society Wolfson Research Merit Award WM170010 and by the Leverhulme Foundation Research Fellowship RF-2020-461\textbackslash \!9. Numerical simulations have been performed on the SUNBIRD system in Swansea, part of the Supercomputing Wales project, which is part-funded by the ERDF via Welsh Government.
\end{acknowledgments}

\appendix
\section{Lattice Parameters}
\label{sec:lattice_parameters}
In Table \ref{tab:lattice_parameters} we provide the simulation parameters used for the production of configurations on the lattices considered in this study. Here $N_{\rm conf}$ indicates the number of thermalized configurations used in the
averages, $am$ is the bare fermion mass in units of the lattice spacing $a$ and the first column is a reference name for the set.
Also indicated for each set is the lattice volume. {\color{black} These data are available in machine-readable format in Ref.~\cite{datapackage}.}

\begin{table}[p]
	\centering
	\caption{The lattices considered in this study. Here $N_{\rm conf}$ indicates the number of thermalized configurations used in the averages, $am$ is the bare fermion mass in units of the lattice spacing $a$ and the first column is a reference name for the set. Also indicated for each set is the lattice volume.}
	\label{tab:lattice_parameters}
	{\resizebox{!}{300pt}{\begin{tabular}{c|cccc}
     & $\beta$ & $am$ & $L_0 \times L_i^3$ & $N_{\rm conf.}$ \\
    \hline
    \hline
    DB1M1 & $2.05$ & $-1.475$ & $32 \times 16^3$ & $4000$ \\
    DB1M1* & $2.05$ & $-1.475$ & $24 \times 12^3$ & $4000$ \\
    DB1M2 & $2.05$ & $-1.49$ & $32 \times 16^3$ & $4000$ \\
    DB1M3 & $2.05$ & $-1.5$ & $24 \times 12^3$ & $4000$ \\
    DB1M4 & $2.05$ & $-1.51$ & $48 \times 24^3$ & $4000$ \\
    DB1M4* & $2.05$ & $-1.51$ & $32 \times 16^3$ & $4163$ \\
    DB1M4** & $2.05$ & $-1.51$ & $24 \times 12^3$ & $4013$ \\
    DB1M5 & $2.05$ & $-1.514$ & $32 \times 16^3$ & $4000$ \\
    DB1M6 & $2.05$ & $-1.519$ & $32 \times 16^3$ & $4000$ \\
    DB1M7 & $2.05$ & $-1.523$ & $48 \times 24^3$ & $4000$ \\
    DB1M7* & $2.05$ & $-1.523$ & $32 \times 16^3$ & $4000$ \\
    DB1M8 & $2.05$ & $-1.524$ & $48 \times 24^3$ & $4000$ \\
    DB1M8* & $2.05$ & $-1.524$ & $32 \times 16^3$ & $4006$ \\
    DB1M9 & $2.05$ & $-1.5246$ & $48 \times 24^3$ & $4000$ \\
    \hline
    DB2M1 & $2.1$ & $-1.42$ & $24 \times 12^3$ & $4000$ \\
    DB2M2 & $2.1$ & $-1.43$ & $24 \times 12^3$ & $4000$ \\
    DB2M3 & $2.1$ & $-1.44$ & $24 \times 12^3$ & $4000$ \\
    DB2M4 & $2.1$ & $-1.45$ & $32 \times 16^3$ & $4000$ \\
    DB2M5 & $2.1$ & $-1.46$ & $32 \times 16^3$ & $4000$ \\
    DB2M6 & $2.1$ & $-1.47$ & $48 \times 24^3$ & $4000$ \\
    DB2M6* & $2.1$ & $-1.47$ & $40 \times 20^3$ & $4000$ \\
    DB2M7 & $2.1$ & $-1.474$ & $64 \times 32^3$ & $4000$ \\
    \hline
    DB3M1 & $2.15$ & $-1.35$ & $24 \times 12^3$ & $4000$ \\
    DB3M2 & $2.15$ & $-1.36$ & $24 \times 12^3$ & $4000$ \\
    DB3M3 & $2.15$ & $-1.37$ & $24 \times 12^3$ & $4000$ \\
    DB3M4 & $2.15$ & $-1.38$ & $32 \times 16^3$ & $4100$ \\
    DB3M5 & $2.15$ & $-1.39$ & $32 \times 16^3$ & $4077$ \\
    DB3M6 & $2.15$ & $-1.4$ & $32 \times 16^3$ & $4014$ \\
    DB3M7 & $2.15$ & $-1.41$ & $48 \times 24^3$ & $4000$ \\
    DB3M8 & $2.15$ & $-1.422$ & $48 \times 24^3$ & $4000$ \\
    \hline
    DB4M1 & $2.2$ & $-1.28$ & $24 \times 12^3$ & $8000$ \\
    DB4M2 & $2.2$ & $-1.29$ & $24 \times 12^3$ & $4320$ \\
    DB4M3 & $2.2$ & $-1.3$ & $32 \times 16^3$ & $4000$ \\
    DB4M3* & $2.2$ & $-1.3$ & $24 \times 12^3$ & $4000$ \\
    DB4M4 & $2.2$ & $-1.31$ & $32 \times 16^3$ & $4000$ \\
    DB4M4* & $2.2$ & $-1.31$ & $24 \times 12^3$ & $4000$ \\
    DB4M5 & $2.2$ & $-1.32$ & $32 \times 16^3$ & $4000$ \\
    DB4M5* & $2.2$ & $-1.32$ & $24 \times 12^3$ & $4000$ \\
    DB4M6 & $2.2$ & $-1.33$ & $24 \times 12^3$ & $4092$ \\
    DB4M7 & $2.2$ & $-1.34$ & $32 \times 16^3$ & $4000$ \\
    DB4M7* & $2.2$ & $-1.34$ & $24 \times 12^3$ & $4000$ \\
    DB4M8 & $2.2$ & $-1.35$ & $32 \times 16^3$ & $4000$ \\
    DB4M9 & $2.2$ & $-1.36$ & $48 \times 24^3$ & $4000$ \\
    DB4M9* & $2.2$ & $-1.36$ & $32 \times 16^3$ & $4000$ \\
    DB4M10 & $2.2$ & $-1.37$ & $48 \times 24^3$ & $4000$ \\
    DB4M11 & $2.2$ & $-1.378$ & $48 \times 24^3$ & $4000$
\end{tabular}
}}
\end{table}

\section{Results}
\label{sec:app_results}
In this appendix, we collect relevant numerical results obtained in the context of this study. Namely, in Table~\ref{tab:baryons} we provide results for baryonic observables, in Table~\ref{tab:mesons} results for mesonic and hybrid (\spinhalf) observables, in Table~\ref{tab:glue} we report glueball masses and the string tension and finally in Table~\ref{tab:topology} we present the topological properties of each ensemble. {\color{black} These data, and the raw data used to generate them, are also available in machine-readable format in Ref.~\cite{datapackage}.}

\begin{table*}
  \centering
  \caption{Baryonic observables.}
  \label{tab:baryons}
  \begin{footnotesize}
  \begin{tabular}{cc|cc|cc|cc}
     & $m_{\mathrm{PCAC}}$ & $am_{2^+_{\mathrm{s}}}$ & $af_{2^+_{\mathrm{s}}}$ & $am_{2^-_{\mathrm{v}}}$ & $af_{2^-_{\mathrm{v}}}$ & $am_{2^-_{\mathrm{ps}}}$ & $af_{2^-_{\mathrm{ps}}}$ \\
    \hline
    \hline
    DB1M1 & $0.14859(18)$ & $0.97087(27)$ & $0.43423(54)$ & --- & --- & --- & --- \\
    DB1M1* & $0.14879(14)$ & $0.97141(32)$ & $0.43443(60)$ & --- & --- & $2.251(12)$ & $0.5919(54)$ \\
    DB1M2 & $0.12838(40)$ & $0.90542(58)$ & $0.4113(14)$ & --- & --- & --- & --- \\
    DB1M3 & $0.11135(20)$ & $0.84527(49)$ & $0.38375(92)$ & --- & --- & --- & --- \\
    DB1M4 & $0.091449(85)$ & $0.76665(15)$ & $0.34838(27)$ & $1.680(46)$ & $0.319(32)$ & $1.7763(85)$ & $0.5197(52)$ \\
    DB1M4* & $0.09041(18)$ & $0.76144(59)$ & $0.3450(12)$ & --- & --- & --- & --- \\
    DB1M4** & $0.09099(33)$ & $0.7644(12)$ & $0.3453(17)$ & --- & --- & --- & --- \\
    DB1M5 & $0.08223(16)$ & $0.72561(60)$ & $0.3278(11)$ & $1.709(22)$ & $0.391(15)$ & $1.635(36)$ & $0.461(27)$ \\
    DB1M6 & $0.06684(21)$ & $0.65234(80)$ & $0.2929(13)$ & $1.538(17)$ & $0.3496(97)$ & --- & --- \\
    DB1M7 & $0.04770(12)$ & $0.54480(42)$ & $0.23919(86)$ & --- & --- & $1.281(10)$ & $0.4119(63)$ \\
    DB1M7* & $0.04947(36)$ & $0.56168(91)$ & $0.24947(92)$ & $1.335(47)$ & $0.302(29)$ & $1.292(23)$ & $0.409(14)$ \\
    DB1M8 & $0.03938(19)$ & $0.49236(62)$ & $0.21206(86)$ & $1.177(17)$ & $0.2606(89)$ & $1.144(18)$ & $0.355(12)$ \\
    DB1M8* & $0.04171(28)$ & $0.5058(16)$ & $0.2190(20)$ & $1.302(13)$ & $0.3247(63)$ & --- & --- \\
    DB1M9 & $0.03425(19)$ & $0.45471(77)$ & $0.1950(12)$ & $1.113(19)$ & $0.2487(96)$ & $1.025(36)$ & $0.309(26)$ \\
    \hline
    DB2M1 & $0.15326(21)$ & $0.97363(51)$ & $0.38982(94)$ & $2.108(12)$ & $0.4685(66)$ & --- & --- \\
    DB2M2 & $0.13820(17)$ & $0.92461(45)$ & $0.37075(71)$ & $1.901(28)$ & $0.377(18)$ & $1.909(20)$ & $0.497(12)$ \\
    DB2M3 & $0.12200(23)$ & $0.86745(64)$ & $0.34875(92)$ & $1.811(23)$ & $0.378(15)$ & --- & --- \\
    DB2M4 & $0.10251(16)$ & $0.79101(51)$ & $0.31313(97)$ & $1.663(15)$ & $0.3554(87)$ & $1.597(23)$ & $0.421(15)$ \\
    DB2M5 & $0.08011(19)$ & $0.69504(63)$ & $0.27437(70)$ & --- & --- & $1.332(31)$ & $0.340(22)$ \\
    DB2M6 & $0.04969(12)$ & $0.53319(48)$ & $0.20450(67)$ & --- & --- & $1.0863(95)$ & $0.3399(62)$ \\
    DB2M6* & $0.05085(15)$ & $0.54108(62)$ & $0.20872(82)$ & $1.119(21)$ & $0.230(12)$ & $1.083(21)$ & $0.329(16)$ \\
    DB2M7 & $0.03246(14)$ & $0.42018(32)$ & $0.15925(36)$ & $0.9406(72)$ & $0.2114(35)$ & $0.833(14)$ & $0.264(12)$ \\
    \hline
    DB3M1 & $0.17742(19)$ & $1.03008(52)$ & $0.36800(86)$ & $1.961(18)$ & $0.369(12)$ & $1.884(26)$ & $0.412(18)$ \\
    DB3M2 & $0.16369(21)$ & $0.98778(60)$ & $0.35295(94)$ & $1.810(94)$ & $0.296(84)$ & --- & --- \\
    DB3M3 & $0.14918(21)$ & $0.94070(72)$ & $0.3360(11)$ & --- & --- & --- & --- \\
    DB3M4 & $0.13407(19)$ & $0.88883(38)$ & $0.31740(51)$ & $1.708(26)$ & $0.342(19)$ & $1.628(12)$ & $0.3969(77)$ \\
    DB3M5 & $0.11737(15)$ & $0.82617(44)$ & $0.29167(75)$ & $1.551(21)$ & $0.293(13)$ & --- & --- \\
    DB3M6 & $0.09851(17)$ & $0.74910(52)$ & $0.26186(81)$ & $1.401(15)$ & $0.2683(86)$ & $1.329(15)$ & $0.334(10)$ \\
    DB3M7 & $0.07808(11)$ & $0.65881(28)$ & $0.22821(33)$ & $1.214(12)$ & $0.2238(68)$ & $1.164(11)$ & $0.3069(78)$ \\
    DB3M8 & $0.04877(11)$ & $0.49990(66)$ & $0.1679(11)$ & $0.971(13)$ & $0.1936(71)$ & $0.8940(97)$ & $0.2643(72)$ \\
    \hline
    DB4M1 & $0.20148(14)$ & $1.07955(38)$ & $0.34586(59)$ & $1.9112(94)$ & $0.3355(54)$ & --- & --- \\
    DB4M2 & $0.18925(23)$ & $1.04361(60)$ & $0.33566(85)$ & $1.792(21)$ & $0.292(13)$ & --- & --- \\
    DB4M3 & $0.17617(14)$ & $1.00319(33)$ & $0.32081(53)$ & $1.754(17)$ & $0.303(11)$ & $1.663(17)$ & $0.338(12)$ \\
    DB4M3* & $0.17601(20)$ & $1.00339(69)$ & $0.3208(11)$ & $1.779(18)$ & $0.316(12)$ & $1.710(30)$ & $0.371(24)$ \\
    DB4M4 & $0.16323(19)$ & $0.96267(35)$ & $0.30796(56)$ & $1.672(30)$ & $0.290(23)$ & $1.577(25)$ & $0.314(20)$ \\
    DB4M4* & $0.16340(20)$ & $0.96441(68)$ & $0.31073(96)$ & $1.599(32)$ & $0.229(19)$ & $1.576(54)$ & $0.314(46)$ \\
    DB4M5 & $0.14881(18)$ & $0.91337(40)$ & $0.29073(65)$ & $1.569(19)$ & $0.262(13)$ & $1.466(24)$ & $0.291(19)$ \\
    DB4M5* & $0.14894(22)$ & $0.91395(78)$ & $0.2915(11)$ & --- & --- & $1.559(11)$ & $0.3622(67)$ \\
    DB4M6 & $0.13422(24)$ & $0.86212(93)$ & $0.2737(12)$ & $1.528(12)$ & $0.2811(71)$ & $1.457(20)$ & $0.344(14)$ \\
    DB4M7 & $0.11902(16)$ & $0.80495(53)$ & $0.25400(76)$ & $1.364(32)$ & $0.216(21)$ & $1.3536(96)$ & $0.3235(66)$ \\
    DB4M7* & $0.11846(26)$ & $0.8032(11)$ & $0.2556(13)$ & $1.306(55)$ & $0.185(37)$ & $1.340(16)$ & $0.325(10)$ \\
    DB4M8 & $0.10217(19)$ & $0.73539(65)$ & $0.22852(96)$ & $1.235(19)$ & $0.197(11)$ & $1.200(14)$ & $0.2820(95)$ \\
    DB4M9 & $0.08438(11)$ & $0.65792(29)$ & $0.20327(36)$ & $1.1663(64)$ & $0.2201(34)$ & $1.0782(82)$ & $0.2740(64)$ \\
    DB4M9* & $0.08437(18)$ & $0.65718(85)$ & $0.2040(11)$ & $1.1573(91)$ & $0.2140(47)$ & --- & --- \\
    DB4M10 & $0.06517(10)$ & $0.56358(39)$ & $0.17199(40)$ & $0.959(13)$ & $0.1609(77)$ & $0.9175(58)$ & $0.2493(39)$ \\
    DB4M11 & $0.04813(13)$ & $0.46875(57)$ & $0.14232(45)$ & $0.790(21)$ & $0.128(13)$ & $0.7555(74)$ & $0.2214(57)$
\end{tabular}

  \end{footnotesize}
\end{table*}

\newpage 

\begin{table}[p]
  \centering
  \caption{Mesonic and hybrid (spin-$\frac{1}{2}$) observables.}
  \label{tab:mesons}
  \begin{tabular}{c|cc|c}
     & $am_{0^-_{\mathrm{v}}}$ & $af_{0^-_{\mathrm{v}}}$ & $am_{\breve{g}}$ \\
    \hline
    \hline
    DB1M1 & $1.15483(45)$ & $0.8696(14)$ & --- \\
    DB1M1* & $1.15624(59)$ & $0.8728(16)$ & --- \\
    DB1M2 & $1.1005(11)$ & $0.8467(33)$ & --- \\
    DB1M3 & $1.0494(12)$ & $0.8107(33)$ & --- \\
    DB1M4 & $0.98420(38)$ & $0.77328(72)$ & --- \\
    DB1M4* & --- & --- & --- \\
    DB1M4** & $0.9770(15)$ & $0.7500(38)$ & --- \\
    DB1M5 & $0.9353(15)$ & $0.7030(50)$ & --- \\
    DB1M6 & $0.8651(15)$ & $0.6395(41)$ & --- \\
    DB1M7 & $0.7491(16)$ & $0.5253(35)$ & $0.911(38)$ \\
    DB1M7* & $0.7707(21)$ & $0.5625(36)$ & --- \\
    DB1M8 & $0.7043(15)$ & $0.4937(25)$ & $0.869(19)$ \\
    DB1M8* & $0.7056(41)$ & $0.478(10)$ & --- \\
    DB1M9 & $0.6502(36)$ & $0.4288(79)$ & $0.810(17)$ \\
    \hline
    DB2M1 & $1.13767(50)$ & $0.7691(13)$ & --- \\
    DB2M2 & $1.09193(60)$ & $0.7397(15)$ & --- \\
    DB2M3 & $1.03989(71)$ & $0.7076(16)$ & --- \\
    DB2M4 & $0.96485(81)$ & $0.6415(22)$ & --- \\
    DB2M5 & $0.86843(78)$ & $0.5705(17)$ & --- \\
    DB2M6 & $0.6933(14)$ & $0.4201(37)$ & --- \\
    DB2M6* & $0.7038(22)$ & $0.4316(53)$ & $0.794(27)$ \\
    DB2M7 & $0.5694(11)$ & $0.3328(18)$ & $0.632(16)$ \\
    \hline
    DB3M1 & $1.16543(45)$ & $0.6888(11)$ & $1.448(17)$ \\
    DB3M2 & $1.12623(60)$ & $0.6684(15)$ & $1.334(33)$ \\
    DB3M3 & $1.08047(72)$ & $0.6396(17)$ & $1.280(36)$ \\
    DB3M4 & $1.02976(62)$ & $0.6069(16)$ & $1.179(35)$ \\
    DB3M5 & $0.96770(86)$ & $0.5620(22)$ & $1.093(29)$ \\
    DB3M6 & $0.89324(95)$ & $0.5180(20)$ & $0.985(24)$ \\
    DB3M7 & $0.80116(48)$ & $0.46144(81)$ & $0.865(26)$ \\
    DB3M8 & $0.6244(30)$ & $0.3274(86)$ & $0.669(12)$ \\
    \hline
    DB4M1 & $1.18981(50)$ & $0.6140(13)$ & $1.311(48)$ \\
    DB4M2 & $1.15655(56)$ & $0.6007(14)$ & $1.217(54)$ \\
    DB4M3 & $1.11683(48)$ & $0.5775(11)$ & $1.256(49)$ \\
    DB4M3* & $1.11830(62)$ & $0.5813(14)$ & --- \\
    DB4M4 & $1.07888(48)$ & $0.5613(11)$ & $1.194(21)$ \\
    DB4M4* & $1.08156(62)$ & $0.5680(13)$ & --- \\
    DB4M5 & $1.02962(46)$ & $0.5317(11)$ & $1.108(44)$ \\
    DB4M5* & $1.03255(77)$ & $0.5378(17)$ & $1.122(42)$ \\
    DB4M6 & --- & --- & $1.071(35)$ \\
    DB4M7 & $0.92276(75)$ & $0.4728(16)$ & $0.904(60)$ \\
    DB4M7* & $0.9179(12)$ & $0.4695(20)$ & --- \\
    DB4M8 & $0.85055(80)$ & $0.4269(17)$ & $0.847(37)$ \\
    DB4M9 & $0.76705(56)$ & $0.3750(11)$ & $0.796(16)$ \\
    DB4M9* & $0.76951(93)$ & $0.3824(16)$ & --- \\
    DB4M10 & $0.66803(51)$ & $0.32129(82)$ & $0.666(13)$ \\
    DB4M11 & $0.56604(97)$ & $0.2681(14)$ & $0.556(12)$
\end{tabular}

  \vspace{0.5cm}
\end{table}

\newpage 

\begin{table}[p!]
	\centering
	\caption{Glueball masses and string tension}
	\label{tab:glue}

    \begin{tabular}{cc|cc}
     & $a \sqrt{\sigma}$ & $am_{0^{++}}$ & $am_{2^{++}}$ \\
    \hline
    \hline
    DB1M1 & --- & $0.929(58)$ & --- \\
    DB1M1* & $0.378(19)$ & $0.96(12)$ & --- \\
    DB1M2 & $0.361(21)$ & $0.834(41)$ & --- \\
    DB1M3 & $0.342(13)$ & $0.730(74)$ & --- \\
    DB1M4 & --- & $0.461(55)$ & --- \\
    DB1M4* & $0.3092(79)$ & $0.577(46)$ & $1.62(13)$ \\
    DB1M4** & $0.3219(74)$ & $0.589(77)$ & --- \\
    DB1M5 & $0.3075(80)$ & $0.554(39)$ & $1.60(15)$ \\
    DB1M6 & $0.2829(49)$ & $0.369(42)$ & $1.78(15)$ \\
    DB1M7 & $0.2391(58)$ & $0.352(24)$ & --- \\
    DB1M7* & $0.2369(83)$ & $0.332(32)$ & --- \\
    DB1M8 & $0.2209(50)$ & $0.253(34)$ & $1.324(93)$ \\
    DB1M8* & $0.2174(53)$ & $0.338(35)$ & $1.25(10)$ \\
    DB1M9 & $0.2072(53)$ & $0.197(17)$ & $1.211(38)$ \\
    \hline
    DB2M1 & $0.3331(50)$ & $0.847(34)$ & $1.73(23)$ \\
    DB2M2 & $0.2987(91)$ & $0.716(57)$ & $1.89(21)$ \\
    DB2M3 & $0.299(11)$ & $0.690(38)$ & $1.55(10)$ \\
    DB2M4 & $0.2719(52)$ & $0.565(36)$ & --- \\
    DB2M5 & $0.2404(69)$ & $0.469(33)$ & --- \\
    DB2M6 & $0.1918(43)$ & $0.373(29)$ & $0.999(83)$ \\
    DB2M6* & $0.1930(31)$ & $0.386(26)$ & $0.902(64)$ \\
    DB2M7 & $0.1536(30)$ & $0.295(13)$ & $0.833(34)$ \\
    \hline
    DB3M1 & $0.3042(35)$ & $0.793(44)$ & $1.62(12)$ \\
    DB3M2 & $0.2939(69)$ & $0.848(43)$ & --- \\
    DB3M3 & $0.2715(40)$ & $0.729(48)$ & --- \\
    DB3M4 & $0.2548(93)$ & $0.722(48)$ & $1.543(91)$ \\
    DB3M5 & $0.2296(67)$ & $0.641(33)$ & $0.94(11)$ \\
    DB3M6 & $0.2145(64)$ & $0.584(18)$ & $1.157(51)$ \\
    DB3M7 & $0.1877(36)$ & $0.541(36)$ & $1.089(43)$ \\
    DB3M8 & $0.1486(24)$ & $0.359(37)$ & $0.818(46)$ \\
    \hline
    DB4M1 & $0.2705(50)$ & $0.761(54)$ & $1.453(85)$ \\
    DB4M2 & $0.2582(32)$ & $0.768(39)$ & $1.364(61)$ \\
    DB4M3 & $0.2479(64)$ & $0.659(78)$ & $1.251(62)$ \\
    DB4M3* & $0.2444(63)$ & $0.800(38)$ & $1.294(74)$ \\
    DB4M4 & $0.2428(67)$ & $0.754(80)$ & $1.235(71)$ \\
    DB4M4* & $0.2346(77)$ & $0.760(42)$ & $1.235(71)$ \\
    DB4M5 & $0.2247(30)$ & $0.700(33)$ & $1.169(50)$ \\
    DB4M5* & $0.2282(39)$ & $0.734(41)$ & $1.28(10)$ \\
    DB4M6 & $0.2091(65)$ & $0.660(30)$ & --- \\
    DB4M7 & $0.1950(68)$ & $0.680(31)$ & $1.134(81)$ \\
    DB4M7* & $0.1919(53)$ & $0.585(31)$ & $1.053(80)$ \\
    DB4M8 & $0.1736(53)$ & $0.563(43)$ & $0.986(45)$ \\
    DB4M9 & $0.168(30)$ & $0.488(70)$ & $0.880(57)$ \\
    DB4M9* & $0.1598(39)$ & $0.485(29)$ & $0.774(70)$ \\
    DB4M10 & $0.1389(32)$ & $0.476(25)$ & $0.761(59)$ \\
    DB4M11 & $0.1202(34)$ & $0.337(31)$ & $0.665(45)$
\end{tabular}

\vspace{0.5cm}
\end{table}

\begin{table}[p!]
  \centering
  \caption{Properties of the topological charge distribution of each ensemble. $Q_0$ denotes the average value of the topological charge $Q$ and $\sigma_Q$ the width of the distribution of $Q$.}
  \label{tab:topology}

  \begin{tabular}{c|cc|c}
     & $Q_0$ & $\sigma_{Q}$ & $\tau_{\exp}$ \\
    \hline
    \hline
    DB1M1 & $-0.36(10)$ & $6.318(75)$ & $2.03(19)$ \\
    DB1M1* & $-0.124(61)$ & $3.766(44)$ & $1.018(57)$ \\
    DB1M2 & $0.01(10)$ & $6.273(74)$ & $2.18(14)$ \\
    DB1M3 & $-0.014(57)$ & $3.549(41)$ & $1.394(70)$ \\
    DB1M4 & $0.04(18)$ & $11.02(13)$ & $1.397(88)$ \\
    DB1M4* & $0.034(88)$ & $5.475(65)$ & $3.06(28)$ \\
    DB1M4** & $0.108(51)$ & $3.207(37)$ & $2.98(23)$ \\
    DB1M5 & $0.045(82)$ & $5.065(59)$ & $2.68(22)$ \\
    DB1M6 & $0.139(78)$ & $4.839(57)$ & $3.69(28)$ \\
    DB1M7 & $-0.02(13)$ & $8.22(10)$ & $1.80(10)$ \\
    DB1M7* & $0.065(65)$ & $4.037(47)$ & $1.88(15)$ \\
    DB1M8 & $-0.20(12)$ & $7.592(89)$ & $1.84(14)$ \\
    DB1M8* & $0.076(60)$ & $3.717(44)$ & $4.39(42)$ \\
    DB1M9 & $0.17(12)$ & $7.258(87)$ & $0.993(39)$ \\
    \hline
    DB2M1 & $0.027(52)$ & $3.240(38)$ & $1.554(74)$ \\
    DB2M2 & $-0.192(50)$ & $3.078(35)$ & $1.554(72)$ \\
    DB2M3 & $0.001(47)$ & $2.950(33)$ & $1.91(12)$ \\
    DB2M4 & $0.034(75)$ & $4.617(55)$ & $2.10(14)$ \\
    DB2M5 & $0.021(65)$ & $4.018(47)$ & $2.09(16)$ \\
    DB2M6 & $-0.43(11)$ & $6.501(81)$ & $3.29(27)$ \\
    DB2M6* & $-0.238(74)$ & $4.574(51)$ & $2.84(21)$ \\
    DB2M7 & $-0.22(14)$ & $8.338(97)$ & $2.05(14)$ \\
    \hline
    DB3M1 & $0.139(46)$ & $2.890(33)$ & $2.04(15)$ \\
    DB3M2 & $-0.033(46)$ & $2.827(32)$ & $2.41(11)$ \\
    DB3M3 & $0.031(42)$ & $2.612(30)$ & $2.52(16)$ \\
    DB3M4 & $0.201(66)$ & $4.099(48)$ & $2.34(20)$ \\
    DB3M5 & $0.196(63)$ & $3.888(45)$ & $2.36(15)$ \\
    DB3M6 & $0.044(56)$ & $3.470(41)$ & $2.75(15)$ \\
    DB3M7 & $-0.13(10)$ & $6.299(76)$ & $3.53(22)$ \\
    DB3M8 & $-0.048(75)$ & $4.609(54)$ & $5.13(34)$ \\
    \hline
    DB4M1 & $-0.015(28)$ & $2.482(20)$ & $2.71(18)$ \\
    DB4M2 & $0.094(37)$ & $2.315(25)$ & $2.70(16)$ \\
    DB4M3 & $0.124(64)$ & $3.982(46)$ & $3.13(23)$ \\
    DB4M3* & $-0.113(35)$ & $2.187(25)$ & $2.74(14)$ \\
    DB4M4 & $-0.089(58)$ & $3.585(42)$ & $3.30(27)$ \\
    DB4M4* & $-0.008(37)$ & $2.281(26)$ & $3.69(18)$ \\
    DB4M5 & $0.077(55)$ & $3.402(40)$ & $3.65(28)$ \\
    DB4M5* & $0.105(34)$ & $2.107(23)$ & $4.63(33)$ \\
    DB4M6 & $0.040(30)$ & $1.870(20)$ & $3.80(26)$ \\
    DB4M7 & $0.048(46)$ & $2.881(32)$ & $3.98(24)$ \\
    DB4M7* & $-0.147(29)$ & $1.801(20)$ & $4.56(24)$ \\
    DB4M8 & $-0.009(43)$ & $2.667(30)$ & $4.93(30)$ \\
    DB4M9 & $-0.145(79)$ & $4.888(57)$ & $6.28(35)$ \\
    DB4M9* & $-0.248(36)$ & $2.256(26)$ & $5.70(35)$ \\
    DB4M10 & $-0.124(64)$ & $3.908(44)$ & $6.42(35)$ \\
    DB4M11 & $-0.358(48)$ & $2.958(35)$ & $7.97(49)$
\end{tabular}

  \vspace{0.5cm}
\end{table}

\begin{table}[p!]
  \centering
  \caption{The $t_0$ and $w_0$ scales for each ensemble, in lattice units. The superscript ``plaq." and ``sym'' refer to the definitions of the density $E$ from the single plaquette and from the symmetrised four-plaquette clover operator respectively. $w_0^{\mathrm{sym.}}$ here is the scale used as $w_0$ elsewhere in the text.}
  \label{tab:flowscales}
  
  \begin{tabular}{c|cc|cc}
     & $\sqrt{8t_0^{\mathrm{plaq.}}}/a$ & $\sqrt{8t_0^{\mathrm{sym.}}}/a$ & $w_0^{\mathrm{plaq.}}/a$ & $w_0^{\mathrm{sym.}}/a$ \\
    \hline
    \hline
    DB1M1 & $2.21398(51)$ & $2.80942(51)$ & $0.88342(34)$ & $0.90605(28)$ \\
    DB1M1* & $2.21564(69)$ & $2.81033(68)$ & $0.88450(43)$ & $0.90688(42)$ \\
    DB1M2 & $2.34765(75)$ & $2.92049(70)$ & $0.93640(41)$ & $0.95515(35)$ \\
    DB1M3 & $2.5089(20)$ & $3.0526(16)$ & $0.99676(85)$ & $1.01351(67)$ \\
    DB1M4 & $2.73611(53)$ & $3.23634(51)$ & $1.07282(30)$ & $1.09027(26)$ \\
    DB1M4* & $2.7457(18)$ & $3.2445(16)$ & $1.07669(84)$ & $1.09406(79)$ \\
    DB1M4** & $2.7202(32)$ & $3.2240(28)$ & $1.0693(15)$ & $1.0863(14)$ \\
    DB1M5 & $2.8665(20)$ & $3.3418(18)$ & $1.11523(82)$ & $1.13312(96)$ \\
    DB1M6 & $3.1344(36)$ & $3.5664(31)$ & $1.2028(13)$ & $1.2220(13)$ \\
    DB1M7 & $3.6132(17)$ & $3.9879(15)$ & $1.36604(68)$ & $1.38643(63)$ \\
    DB1M7* & $3.5669(42)$ & $3.9462(34)$ & $1.3514(16)$ & $1.3712(16)$ \\
    DB1M8 & $3.8718(23)$ & $4.2228(21)$ & $1.45400(86)$ & $1.47563(83)$ \\
    DB1M8* & $3.8220(66)$ & $4.1766(65)$ & $1.4392(26)$ & $1.4596(27)$ \\
    DB1M9 & $4.0656(22)$ & $4.3985(20)$ & $1.51666(81)$ & $1.53894(89)$ \\
    \hline
    DB2M1 & $2.5678(19)$ & $3.0749(16)$ & $1.03867(87)$ & $1.05742(76)$ \\
    DB2M2 & $2.7229(23)$ & $3.1988(20)$ & $1.0873(10)$ & $1.10611(93)$ \\
    DB2M3 & $2.9105(28)$ & $3.3536(22)$ & $1.1458(12)$ & $1.1659(11)$ \\
    DB2M4 & $3.2950(21)$ & $3.6840(19)$ & $1.23910(95)$ & $1.25959(88)$ \\
    DB2M5 & $3.7156(37)$ & $4.0556(34)$ & $1.3760(15)$ & $1.3977(16)$ \\
    DB2M6 & $4.6124(30)$ & $4.8842(33)$ & $1.6714(13)$ & $1.6956(15)$ \\
    DB2M6* & $4.5505(39)$ & $4.8252(41)$ & $1.6496(17)$ & $1.6732(16)$ \\
    DB2M7 & $5.4980(31)$ & $5.7318(31)$ & $1.9644(12)$ & $1.9917(12)$ \\
    \hline
    DB3M1 & $3.0042(26)$ & $3.4138(27)$ & $1.1608(13)$ & $1.1820(13)$ \\
    DB3M2 & $3.1483(34)$ & $3.5364(31)$ & $1.2058(14)$ & $1.2273(15)$ \\
    DB3M3 & $3.3224(43)$ & $3.6910(40)$ & $1.2657(19)$ & $1.2871(19)$ \\
    DB3M4 & $3.6002(29)$ & $3.9360(27)$ & $1.3225(12)$ & $1.3451(10)$ \\
    DB3M5 & $3.8580(31)$ & $4.1673(31)$ & $1.4070(14)$ & $1.4291(13)$ \\
    DB3M6 & $4.2426(43)$ & $4.5212(43)$ & $1.5320(18)$ & $1.5549(17)$ \\
    DB3M7 & $4.7560(34)$ & $5.0023(28)$ & $1.6987(13)$ & $1.7227(12)$ \\
    DB3M8 & $5.9555(75)$ & $6.1593(81)$ & $2.0940(32)$ & $2.1213(29)$ \\
    \hline
    DB4M1 & $3.5912(30)$ & $3.9135(30)$ & $1.3071(14)$ & $1.3295(14)$ \\
    DB4M2 & $3.7187(50)$ & $4.0295(51)$ & $1.3511(22)$ & $1.3737(23)$ \\
    DB4M3 & $3.8608(32)$ & $4.1583(29)$ & $1.3951(14)$ & $1.4183(14)$ \\
    DB4M3* & $3.8733(50)$ & $4.1690(51)$ & $1.4026(25)$ & $1.4247(22)$ \\
    DB4M4 & $4.0069(36)$ & $4.2904(36)$ & $1.4400(15)$ & $1.4638(16)$ \\
    DB4M4* & $4.0230(66)$ & $4.3061(65)$ & $1.4512(31)$ & $1.4743(32)$ \\
    DB4M5 & $4.2441(44)$ & $4.5096(39)$ & $1.5203(19)$ & $1.5432(17)$ \\
    DB4M5* & $4.2569(78)$ & $4.5225(83)$ & $1.5277(43)$ & $1.5509(40)$ \\
    DB4M6 & $4.507(11)$ & $4.759(11)$ & $1.6153(51)$ & $1.6392(45)$ \\
    DB4M7 & $4.7605(66)$ & $4.9929(58)$ & $1.6868(25)$ & $1.7098(25)$ \\
    DB4M7* & $4.836(13)$ & $5.070(13)$ & $1.7279(56)$ & $1.7517(66)$ \\
    DB4M8 & $5.2149(89)$ & $5.4324(85)$ & $1.8448(37)$ & $1.8692(35)$ \\
    DB4M9 & $5.7286(60)$ & $5.9249(56)$ & $2.0061(25)$ & $2.0319(28)$ \\
    DB4M9* & $5.819(14)$ & $6.013(15)$ & $2.0516(66)$ & $2.0759(67)$ \\
    DB4M10 & $6.5742(97)$ & $6.7521(96)$ & $2.2901(41)$ & $2.3159(39)$ \\
    DB4M11 & $7.728(18)$ & $7.891(19)$ & $2.6863(78)$ & $2.7139(75)$
\end{tabular}

\end{table}

\pagebreak

\bibliographystyle{apsrev4-2.bst}
\bibliography{references}

\end{document}